\documentclass[reprint,superscriptaddress,amsmath,amssymb,aps,prb,longbibliography]{revtex4-2}
\usepackage[utf8]{inputenc}
\usepackage{graphicx}
\usepackage{dcolumn}
\usepackage{bm}
\usepackage[colorlinks,citecolor=blue,linkcolor=blue,urlcolor=blue]{hyperref}
\usepackage[normalem]{ulem}

\newcommand{\zhang}[1]{{\color{black}{#1}}}

\begin{document}

\title{Pairing phase diagram for electron-doped cuprates in the square-lattice $t-U-V$ Hubbard model}

\author{Zhangkai Cao}
\thanks{These authors contributed equally.}
\affiliation{School of Science, Harbin Institute of Technology, Shenzhen, 518055, China}

\author{Shuning Tan}
\thanks{These authors contributed equally.}
\affiliation{Key Laboratory for Microstructural Material Physics of Hebei Province, School of Science, Yanshan University, Qinhuangdao 066004, China} 

\author{Ji Liu}
\affiliation{School of Science, Harbin Institute of Technology, Shenzhen, 518055, China}
\affiliation{Shenzhen Key Laboratory of Advanced Functional Carbon Materials Research and Comprehensive Application, Shenzhen 518055, China.}

\author{Xiaosen Yang}
\affiliation{Department of Physics, Jiangsu University, Zhenjiang, 212013, China}

\author{Tao Ying}
\email{taoying86@hit.edu.cn}
\affiliation{School of Physics, Harbin Institute of Technology, Harbin 150001, China}

\author{Ho-Kin Tang}
\email{denghaojian@hit.edu.cn}
\affiliation{School of Science, Harbin Institute of Technology, Shenzhen, 518055, China}
\affiliation{Shenzhen Key Laboratory of Advanced Functional Carbon Materials Research and Comprehensive Application, Shenzhen 518055, China.}

\author{Cho-Tung Yip}
\affiliation{School of Science, Harbin Institute of Technology, Shenzhen, 518055, China}
 
\date{\today}

\begin{abstract} 
Motivated by significant discrepancies between experimental observations of electron-doped cuprates and numerical results of the Hubbard model, we investigate the role of nearest-neighbor (NN) electron interactions $V$ by studying the $t-U-V$ model on square lattices. Upon doping $\delta$= 0.153, by using constrained path quantum Monte Carlo (CPQMC) method, we find that NN electron attraction $V$ can notably drive an exotic $p$-wave spin-triplet pairing, while the NN electron repulsion $V$ will suppress the $d_{x^2-y^2}$-wave ($d$-wave) pairing and triggers the $d_{xy}$-wave pairing. 
Especially in the intermediate coupling regime, as NN repulsion  increases, the intensity of $d_{xy}$-wave pairing also increases, further suppressing the presence of $d$-wave pairing, which may help explain the notable suppression of $d$-wave pairing in electron-doped cuprate superconductors.
Besides the pairing phase, we also find that the NN electron attraction $V$ has no significant effect on spin density wave (SDW) and charge density wave (CDW), but repulsion $V$ significantly enhanced CDW and suppressed SDW.
Our study suggests the $t-U-V$ Hubbard model can serve as the minimal model to capture the essential physics of the electron-doped cuprates. 

\end{abstract} 
\maketitle

\section{Introduction}

In almost four decades since the discovery of cuprate high-temperature ($T_c$) superconductors \cite{Bednorz1986}—later identified with a $d_{x^2-y^2}$-wave ($d$-wave) pairing symmetry \cite{Tsuei2000-fb}, understanding the pairing mechanism of high-$T_c$ superconductivity (SC) as one of the great challenges in modern condensed matter physics remains substantially enigma\cite{Imada1998-fh,Orenstein2000-am,Lee2006-qq,Armitage2010-kn,Fradkin2015-xz,Zhou2021-xg, Keimer2015-jp}. Theoretically, this foundational puzzle was cast within the framework of the two-dimensional (2D) Hubbard or $t-J$ model and their variants \cite{Keimer2015-jp,Zhang1988-dw,Anderson2004-gq,Jiang2021-hj,Qin2022-ad}, the strong electron repulsive interactions in 3d orbitals are believed to play crucial roles \cite{Mai2021-wv}. Specifically, experimental findings reveal a notable $d$-wave SC feature on the hole-doped side \cite{Lee2006-qq,Scalapino2012-jr, Keimer2015-jp,Sobota2021-du}, whereas the search for a $d$-wave SC remains elusive in numerical simulations \cite{Jiang2019-km,qin2020absence,Chen2023-rk,Xu2024-wt,Lu2024-ah}. Moreover, on the electron-doped side, theoretically people find a strong $d$-wave SC over a broader doping range than the phase diagram of the experiment \cite{Motoyama2007-th,Scalapino2012-jr,Jiang2021-hj,Chen2023-rk,Xu2024-wt,Sobota2021-du}. Therefore, such theoretical exploration of electron and hole doping requires studying single-band Hubbard or $t-J$ models with different additional ingredients like next-nearest-neighbor (NNN) hopping $t^{\prime}$ \cite{Jiang2019-km,Jiang2021-hj,Xu2024-wt} or nearest-neighbor (NN) interaction $V$ \cite{Chen2024-rr}, so that these models can more precisely and simultaneously capture the qualitative physics of the corresponding real cuprate materials.

Recently, dome-like $d$-wave SC region is found in both the electron-doped ($t^{\prime}=0.2t$) and hole-doped ($t^{\prime}=-0.2t$) regimes of the 2D $t-t^{\prime}-U$ Hubbard model \cite{Xu2024-wt}, a nonzero $t^{\prime}$ is necessary to account for the particle-hole asymmetry and $d$-wave SC \cite{qin2020absence}. 
Both positive and negative results have been observed for $d$-wave pairing order through various numerical simulation methods, as well as the coexistence and competition between $d$-wave SC, antiferromagnetic order (AF) \cite{Uefuji2001-vc}, charge density wave (CDW) \cite{Comin2014-bz}, spin density wave (SDW) \cite{Moon2009-il}, pair density wave (PDW) \cite{Shi2020-vh}, electron nematic order \cite{Sato2017-bt} and pseudogap (PG) \cite{Doiron-Leyraud2017-lv,He2014-bx}, reflecting the Hubbard model and its cousin $t-J$ model are extreme sensitivity to ground state configurations and low-lying excitations \cite{Himeda2002-mc,Jiang2021-hj,Chen2024-rr,Xu2024-wt,Corboz2014-kr,Zheng2016-xt,qin2020absence}.
Moreover, recent angle-resolved photoemission spectroscopy (ARPES) experiments on one-dimensional (1D) cuprate chains Ba$_{2-x}$Sr$_x$CuO$_{3+\delta}$ \cite{Chen2021-xx} have revealed an anomalously sizable attractive interaction between NN electrons, possibly mediated by phonons \cite{Chen2021-xx,Wang2021-vs,Tang2023-aa}. Such an effective attraction, although not so strong as the on-site Coulomb repulsion, is comparable to the electron hopping integral ($V\sim-t$) \cite{Chen2021-xx}, and thereby also should not be ignored in 2D systems \cite{Wang2021-vs,Tang2023-aa,Peng2023-on,Cao2024-hj,Lu2024-bh}. Furthermore, many recent studies have amassed evidence highlighting the crucial role of NN repulsion in both hole- and electron-doped cuprates \cite{Misawa2014-at,Sau2014-pc,Jiang2018-sc,Hirayama2018-tm,Banerjee2022-vp, Bejas2022-gk, Boschini2021-xf,Chen2024-rr,PhysRevB.87.075123,PhysRevB.94.155146}, which induced by the nonlocal Coulomb repulsion. 
We consider to be consistent with the amplitude of the anomalous NN interaction $|V|\sim t$ extracted from both experimental and theoretical studies \cite{Chen2021-xx,Wang2021-vs,Tang2023-aa,Jiang2018-sc,Jiang2022-ok,Peng2023-on,Cao2024-hj,Lu2024-bh,Misawa2014-at,Sau2014-pc,Jiang2018-sc,Hirayama2018-tm,Banerjee2022-vp, Bejas2022-gk, Boschini2021-xf,PhysRevB.87.075123,PhysRevB.94.155146}.

\zhang{From an experimental perspective \cite{Sobota2021-du}, there is a basic consensus that the SC region of electron doping is indeed limited to a very small area near doping ~0.15, and the $T_c$ is very low, which is in sharp contrast to hole doping. This asymmetry strongly suggests a fundamental difference in the underlying interactions that govern pairing. In realistic systems, the Coulomb repulsion is hardly screened to a completely local interaction but has a short-ranged nonlocal contribution.
For the cuprates, Sénéchal et al. \cite{PhysRevB.87.075123} and Reymbaut et al. \cite{PhysRevB.94.155146} estimated a NN Coulomb repulsion of ~400 meV. Such a sizeable interaction can suppress pairing tendencies that rely on short-range coherence, especially in more itinerant electron-doped systems where screening is more effective and the influence of spin fluctuations is comparatively reduced. Therefore, we argue that the presence of a significant NN repulsion $V>0$ is a key factor limiting SC in electron-doped cuprates, in contrast to the hole-doped side where strong correlations and spin fluctuations may partially overcome or even reverse this tendency via effective attractive interactions \cite{Chen2021-xx}.
We believe that the NN repulsion $V$ plays a very important role in electron-doped cuprates.}

\zhang{The rest of this paper is outlined as follows.  
Sec.\ \ref{Model} details the Hamiltonian of $t-U-V$ Hubbard model in 2D lattice. 
Our results and discussion are presented in  Sec.\ \ref{Results and discussion}: First is a general phase diagram scan of the electron doping $\delta$= 0.153, we find that NN electron attraction $V$ can notably drive an exotic $p$-wave spin-triplet pairing, while the NN electron repulsion $V$ will suppress the $d$-wave pairing and triggers the $d_{xy}$-wave pairing. Second is the NN repulsion $V$ significantly enhanced CDW and suppressed SDW. Moreover, as doping increases, the dominant pairing region of $p$-wave and $d_{xy}$-wave also expands, further suppressing the presence of $d$-wave. Our study suggests the $t-U-V$ Hubbard model can serve as the minimal model to capture the essential physics of the electron-doped cuprates. Finally, we give a summary in Sec.\ \ref{SUMMARY}. }

\section{Model}
\label{Model}

Motivated by the above studies, we examine the roles of the NN electron interactions in the $t-U-V$ model on the square lattice, aiming at identifying a minimal model capable of describing the phase diagram of the electron-doped cuprates. We want to find the reason why SC is suppressed on the electron-doped side. The Hamiltonian is given by
\begin{equation}\label{eq:hamiltonian}
\begin{aligned}
H =&-\sum_{\substack{i, j, \sigma}} t_{i, j}  \left(c_{i,\sigma}^\dag c_{j,\sigma} + h.c. \right)
+ U \sum_i n_{i,\uparrow} n_{i,\downarrow} \\
&+ V \sum_{\substack{i, j}} n_{i} n_{j}
\end{aligned}
\end{equation}
where $c_{i,\sigma}^\dag$ ($c_{i,\sigma}^{\,}$) is electron creation (annihilation) operator with spin $\sigma$ = $\uparrow,\downarrow$, and $n_{i,\sigma}=c_{i,\sigma}^\dag c_{i,\sigma}$ is the electron number operator. The electron hopping amplitude $t_{i, j} = t$ if $i$ and $j$ are the NN sites, and $t_{i, j} = t^{\prime}$ for NNN sites. Here, we set $t=1$ as the energy unit. We use $t^{\prime}=0.2$ for electron doping, according to band-structure calculations in cuprates \cite{Andersen1995-pi,Hirayama2018-tm}. $U > 0$ is the on-site Coulomb repulsion. $V$ is the NN electron interactions, where $V < 0$ and $V > 0$ represent electron attraction and repulsion, respectively.
In our recent work \cite{Cao2024-hj}, we find that NN electron attraction $V$ significantly promotes an exotic spin-triplet ($p$-wave) pairing phase in the hole doping region. Meanwhile, the $d$-wave pairing in the intermediate coupling regime shows insignificant response to the increase of $V$.
Here, we systematically study the phase diagrams in the $t-U-V$ model for electron doping by using the constrained path quantum Monte Carlo~(CPQMC) \cite{Zhang1995-hn,Zhang1997-mr}, which partially solves sign problem concerning only the ground-state properties of the systems of interacting fermions at zero temperature. 


We define the effective pair momentum distribution function \cite{White1989-fn,Ying2018-ev}
\begin{equation}
N^{\rm eff}_{\mathrm \zeta-pair}({\bf k}) = (1/N)\sum_{i,j} \mbox{exp}[i{\bf k}({\bf r}_i-{\bf r}_j)]C^{\rm eff}_{\mathrm \zeta-pair}(i,j),
\label{nspdtkpair}
\end{equation}
where $\zeta =  d,\ d_{xy},\ p$ and $N$ is the number of lattice sites. The effective real space correlation for $d$-wave and $d_{xy}$-wave pairing operator $C^{\rm eff}_{\zeta-pair}(i,j) = \sum_{\delta_{\zeta},\delta'_{\zeta}}\langle {\Delta}_{\zeta}^{\dagger}(i,i+\delta_{\zeta}) {\Delta}_{\zeta}(j,j+\delta'_{\zeta}) \rangle - G^{\uparrow}_{i,j}G^{\downarrow}_{i+\delta_{\zeta},j+\delta'_{\zeta}})$. For $d$-wave, where $\Delta_{d}^\dagger(i)=c^\dagger_{i\uparrow}(c^\dagger_{i+x \downarrow}-c^\dagger_{i+y \downarrow}+c^\dagger_{i-x \downarrow}-c^\dagger_{i-y \downarrow})$ and $\delta^{(')}_{\zeta}$ are NN sites. For $d_{xy}$-wave, where $\Delta_{d_{xy}}^\dagger(i)=c^\dagger_{i\uparrow}(c^\dagger_{i+x+y \downarrow}-c^\dagger_{i-x+y \downarrow}+c^\dagger_{i-x-y \downarrow}-c^\dagger_{i+x-y \downarrow})$ and $\delta^{(')}_{\zeta}$ are the NNN sites. $G^{\sigma}_{i,j}= \langle c_{i\sigma}c^\dagger_{j\sigma} \rangle$ is the Green's function and $G^{\sigma}_{i,j}G^{\sigma}_{i,j}$ is the uncorrelated pairing structure factor. 
The effective real-space correlation for $p$-wave pairing operator $C^{\rm eff}_{ p-pair}(i,j) = \sum_{\delta_{\zeta},\delta'_{\zeta}}\langle {\Delta}_{p}^{\dagger}(i,i+\delta_{\zeta}) {\Delta}_{p}(j,j+\delta'_{\zeta}) \rangle - G^{\sigma}_{i,j}G^{\sigma}_{i+\delta_{\zeta},j+\delta'_{\zeta}})$, in which $\delta^{(')}_{\zeta}$ are the NN sites. $\Delta_{p}^\dagger(i)=c^\dagger_{i\sigma}(c^\dagger_{i+l \sigma}-c^\dagger_{i-l \sigma})$, with spin $\sigma=\uparrow, \downarrow$ representing the $\uparrow \uparrow$ and $\downarrow \downarrow$ pairing, and $l = x,\ y$ corresponds to the symmetries of $p_x$ and $p_y$, respectively. 


\begin{figure}[!]
    \centering 
    \includegraphics[width=1\linewidth]{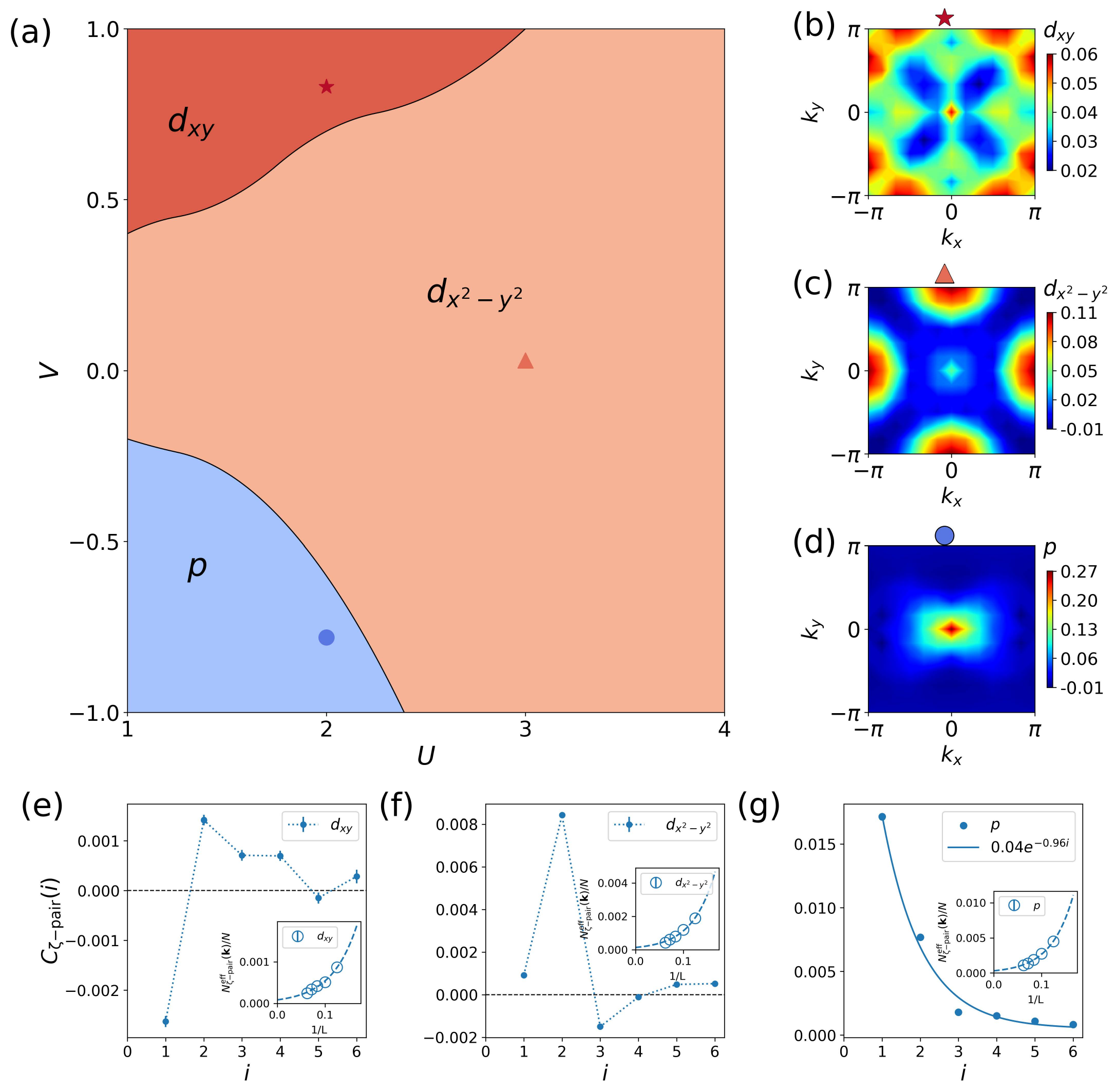}
    \caption{(Color online) Phase diagram in $t-U-V$ Hubbard model. (a) Schematic zero-temperature phase diagram at $\delta$ = 0.153 for electron doping on 12 $\times$ 12 lattice, there are $d_{xy}$-wave, $d$-wave and exotic $p$-wave triplet pairing phases. (b) CPQMC simulation result of effective $d_{xy}$-wave pair momentum distribution function $N^{\rm eff}_{d_{xy}-pair}({\bf k})$ with $U= 2$ and $V= 0.8$. (c) $N^{\rm eff}_{d-pair}({\bf k})$ with $U= 3$ and $V= 0.0$. (d) $N^{\rm eff}_{p-pair}({\bf k})$ with $U= 2$ and $V= -0.8$. (e)-(g) The effective real space correlation $C^{\rm eff}_{\zeta-pair}$ for different pairing mode consistent with (b)-(d). The insets of (e)-(g) show the lattice size effect for different pairing mode consistent with (b)-(d), fitted using exponential function in $1/L$.
    }
    \label{fig1}
\end{figure}

\section{Results and discussion}  
\label{Results and discussion}

We focus on the optimal doping $\delta$ = 0.153 \cite{Keimer2015-jp}. 
At zero temperature, based on the CPQMC simulation of the $t-U-V$ model, we summarize our main findings in the phase diagram, where the $d_{xy}$-wave, $d$-wave and exotic $p$-wave pairing phases in the ground state is uncovered, as depicted in Fig.\ \ref{fig1}(a).
In the absence of NN electronic interaction ($V = 0$), the $d$-wave pairing phase dominates the system. However, with the emergence of the NN electronic attraction $V$, even a slight enhancement of $V$ can propel the system into a $p$-wave triplet pairing phase, particularly for small $U$. Conversely, NN electronic repulsion ($V > 0$) favors the formation of $d_{xy}$-wave pairing phase, while suppressing the $d$-wave pairing.

Then we analyze the characteristics of three types effective pairing correlation function in momentum space, as shown in Figs.\ \ref{fig1}(b)-(d). 
Both $d_{xy}$-wave and $p$-wave pairing exhibit condensation at zero momentum ($Q=(0,0)$).
Interestingly, in addition to a strong condensation at zero momentum, the $d_{xy}$-wave pairing also exhibits eight relatively weak peaks near the $(\pi, \pi)$ point, while preserving the four-fold rotational ($C_4$) symmetry. In contrast, the $p$-wave pairing is only condenses at point $Q=(0,\ 0)$, corresponding to weak-coupling $U$ and strong attraction $V$ region, supporting the formation of Cooper-pair triplets with zero center-of-mass momentum and $p$-wave symmetry. The $d$-wave pairing exhibits a singular nonzero condensation structure, and its maximum value occurs near nonzero momentum point $Q=(\pi,0)$, which represents the uneven distribution of $d$-wave pairing in momentum space. \zhang{From another perspective, the $d$-wave pairing suggests a $d$-wave PDW $(\pi,0)$ state.}

Meanwhile, the real space pairing correlation functions further elucidate the distinct characteristics of $d_{xy}$-wave, $d$-wave, and $p$-wave pairings, as shown in Figs.\ \ref{fig1}(e)-(g). The $d_{xy}$-wave pairing exhibits a slightly staggered behavior, consistent with its have incommensurate condensation points near $(\pi,\ \pi)$.  
The $d$-wave pairing shows staggered behavior depending on pairing directions. Conversely, the $p$-wave pairing shows exponential decay in real space and drops to 0 quickly in the distance, indicating short-range correlations.
The insets of Figs.\ \ref{fig1}(e)-(g), we show the CPQMC results for the lattice size effect for different pairing mode consistent with Figs.\ \ref{fig1}(b)-(d). The finite-size extrapolation for various pairing modes remains positive under different conditions, suggesting that these pairing states are robust.

\begin{figure}[!]
    \centering 
    \includegraphics[width=1\linewidth]{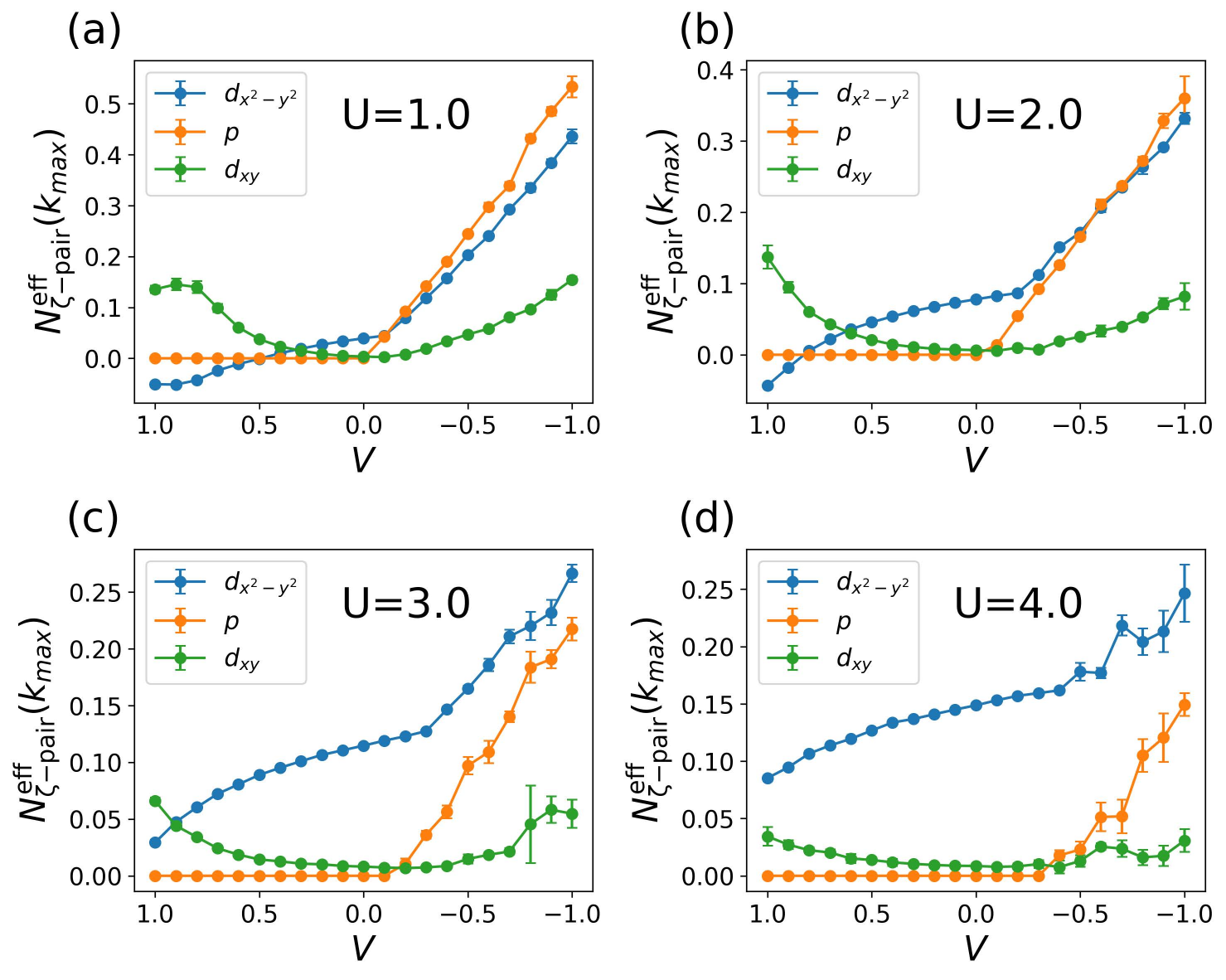}
    \caption{(Color online) The competition and distinction of $d_{xy}$-wave, $d$-wave and $p$-wave. (a)-(d) CPQMC simulation results of the effective pair momentum distribution function $N^{\rm eff}_{d-pair}({\bf k}_{\rm max})$, $N^{\rm eff}_{d_{xy}-pair}({\bf k}_{\rm max})$ and $N^{\rm eff}_{p-pair}({\bf k}_{\rm max})$ at $\delta$ = 0.153 as a function of $V$ on 12 $\times$ 12 lattice, and fix (a) $U$ = 1.0, (b) $U$ = 2.0, (c) $U$ = 3.0, (d) $U$ = 4.0. ${\bf k}_{\rm max}$ is the momentum where the maximum value in momentum space.   }
    \label{fig2}
\end{figure}

To clearly illustrate the competitive interplay and distinct characteristics of the effective $d$-wave, $d_{xy}$-wave and $p$-wave pairing phases, we present  simulations traversing the vertical trajectory in the phase diagram at $\delta=$ 0.153. This approach enables a visualization of how these pairing phases differentiate and compete within the phase space, as depicted in Fig.\ \ref{fig2}. Firstly, we found that in the absence of $V$, the dominant pairing in the system is the $d$-wave, and its intensity increases with the increase of $U$. The behavior of the $d$-wave pairing varies significantly depending on $U$ as the NN interaction transitions from attractive ($V<0$) to repulsive ($V>0$). As $V$ varies, the behavior of the $d$-wave pairing strength can be categorized into two regimes: (i) Weak Coupling Regime ($U = 1,\ 2$), the $d$-wave pairing strength rapidly decreases with the $V$ shifts from attraction to repulsion. Under extreme NN repulsion ($V>0$), the $d$-wave pairing strength becomes negative, indicating strong suppression of the $d$-wave phase; (ii) Intermediate Coupling Regime ($U = 3,\ 4$, particularly $U=4$): As $V$ increases, the $d$-wave pairing strength slowly decreases continuously, indicating substantial suppression of the $d$-wave pairing phase under NN repulsion.


In the weak coupling regime, the strength of the $p$-wave pairing increases dramatically with the enhancement of attraction $V$, rapidly surpassing the strength of the $d$-wave pairing. Conversely, in the intermediate coupling regime, the $p$-wave pairing remains suppressed until $V$ attains a critical threshold, whereupon it undergoes a rapid surge. Notably, when $V\geq0$, the $p$-wave pairing is completely absent, highlighting that $p$-wave pairing is exclusively facilitated by NN electron attraction ($V < 0$), while NN repulsion impedes the $p$-wave pairing formation. Furthermore, our analysis explores the behavior of the $d_{xy}$-wave pairing under variations in $V$. Regardless of whether $V$ is attractive or repulsive, increasing $V$ consistently fosters the growth of $d_{xy}$-wave pairing.
This effect is particularly pronounced under strong repulsion $V$, where $d_{xy}$-wave pairing grows significantly and ultimately surpass the $d$-wave pairing. This phenomenon is closely linked to the suppression of $d$-wave pairing observed in electron-doped cuprates.

The intermediate coupling $U$ is believed to more accurately describe the behavior of cuprate superconductors. For intermediate $U$ and attractive $V$, the $d$-wave pairing significantly increase with the increase of $V$, indicating that the $d$-wave pairing is relatively sensitive to attractive $V$ in strongly correlated systems. On the repulsive $V$ side, the $d$-wave pairing significantly decreases as $V$ increases, indicating that $d$-wave pairing is suppressed by the repulsive $V$.
For the overall phase diagram (Fig.\ \ref{fig1}(a)), $d$-wave pairing dominates when $V$ is absent or small. On the attractive $V$ side, the $p$-wave pairing region expands as its strength increases with $V$. On the repulsive $V$ side, the regime dominated by $d_{xy}$-wave pairing gradually increases, due to the enhancement of $d_{xy}$-wave pairing and the suppression of $d$-wave pairing by repulsive $V$. This indicates that repulsive $V$ is unfavorable for $d$-wave pairing, which may lead to the shrinkage of $d$-wave SC region and the decrease of $T_c$ in electron-doped cuprates when the NN repulsion increases.

\begin{figure}[!]
    \centering 
    \includegraphics[width=1\linewidth]{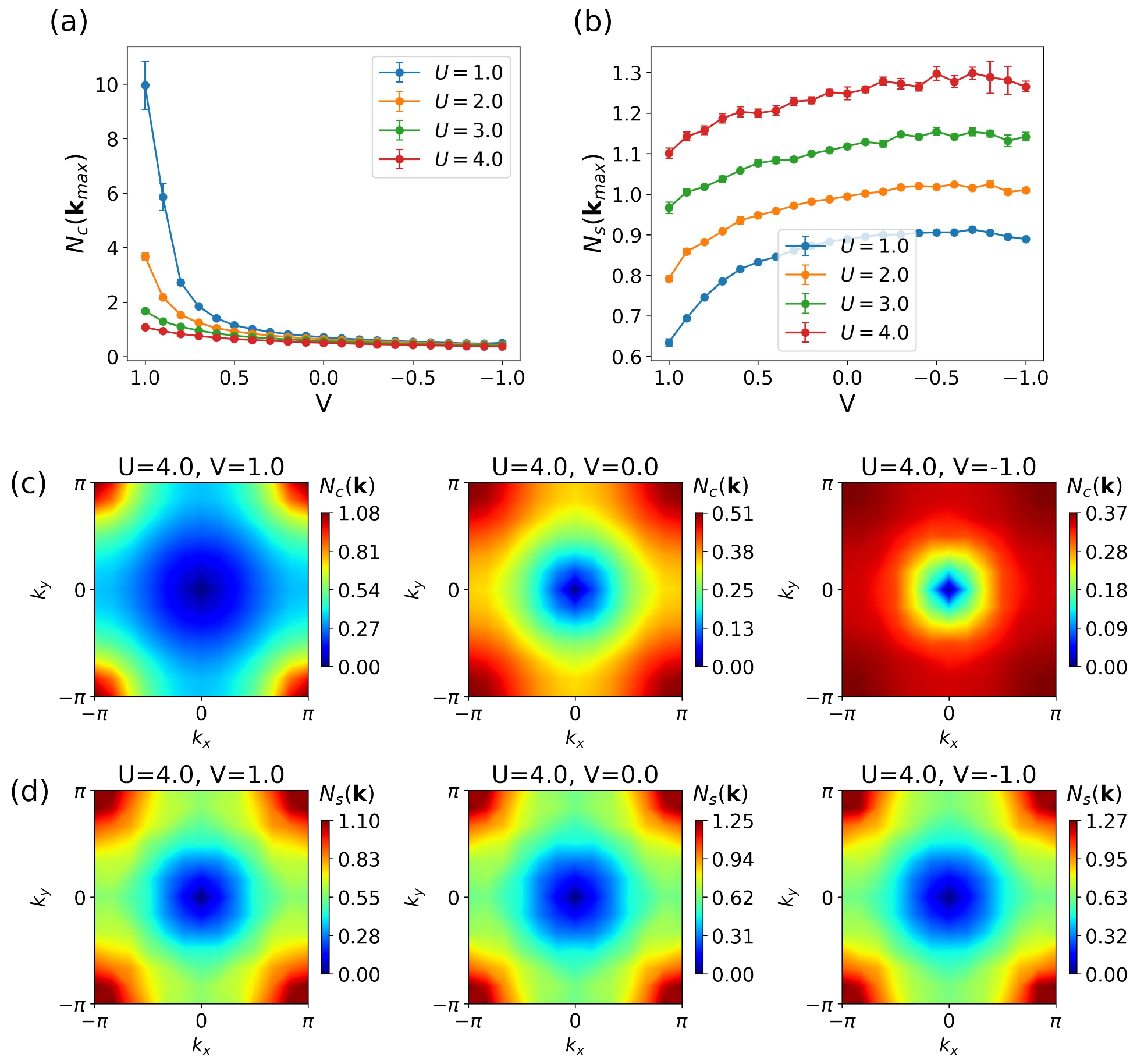}
    \caption{(Color online) Charge and spin density correlation. CPQMC simulation of (a) charge structure factor $N_{c}({\bf k})$ and (b) spin structure factor $N_{s}({\bf k})$ at $\delta$ = 0.153 as a function of $V$ on 12 $\times$ 12 lattice, for $U$ = 1.0, 2.0, 3.0, 4.0. (c) $N_{c}({\bf k})$ and (d) $N_{s}({\bf k})$ at $\delta$ = 0.153 with $U = 4$ and $V= 1.0, 0.0, -1.0$ on 12 $\times$ 12 lattice, respectively. Both $N_{s}({\bf k})$ and $N_{c}({\bf k})$ peak at the antiferromagnetic ordering momentum, ${\bf k}_{\rm max}$ = ($\pi$, $\pi$).}
    \label{fig3}
\end{figure}

To explore the instability apart from superconductivity, we also study the magnetic and charge instabilities, such as charge and spin density waves. We define the charge structure factor and spin structure factor in particle-hole channel, 
\begin{equation}
N_{\mathrm c}({\bf k}) = (1/N)\sum_{i,j} \mbox{exp}[i{\bf k}({\bf r}_i-{\bf r}_j)]\, \langle {n}_i {n}_j \rangle,
\label{nskpair}
\end{equation}
\begin{equation}
N_{\mathrm s}({\bf k}) = (1/N)\sum_{i,j} \mbox{exp}[i{\bf k}({\bf r}_i-{\bf r}_j)]\, \langle {\bf S}_i \cdot{\bf S}_j \rangle,
\label{nskpair}
\end{equation}
Here, the density number operator is defined as
${n}_i = \sum_{\sigma} c^\dagger_{i \sigma} c_{i \sigma}$ and ${\bf S}_i$ the spin operator at site $i$.
The correlation function of charge density wave and spin density wave in real space are defined as $C_{\rm CDW} = \langle {n}_i {n}_j \rangle$ and $C_{\rm SDW} = \langle {\bf S}_i \cdot{\bf S}_j \rangle$.

To elucidate the effect of $V$ on charge and spin density correlation, we present simulations by fixing $U$ = $1,\ 2,\ 3,\ 4$ and varying $V$, as shown in Figs.\ \ref{fig3}(a) and (b).
Additionally, we present the momentum-space distribution maps of CDW and SDW in Figs.\ \ref{fig3}(c) and (d) for $U = 4$ and $V= 1.0, 0.0, -1.0$. As observed, the maximum values of both CDW and SDW occur at ($\pi$, $\pi$), indicating that ${\bf k}_{\rm max}$ = ($\pi$, $\pi$).
Overall, the behavior of charge correlation shows an increasing trend when $V$ shifts from attraction to repulsion, and the increase is more pronounced on the repulsion $V$ side. In contrast, the spin correlation remains basically unchanged on attractive $V$ side, but significantly decreases as $V$ increases on the repulsive $V$ side. Specifically, the charge correlation slowly increases with the decrease of $V$ on the attractive $V$ side, but remains lower than the strength of spin correlation. However, on the repulsive $V$ side, the charge correlation continues to gradually increase as $V$ increases and eventually surpasses the spin correlation when $V$ is strong enough.
This indicates that the system tends to exhibit an SDW order without NN interaction or under attractive $V$, and tends to form a CDW order for large repulsion $V$. In other words, the repulsive $V$ is demonstrated to suppress the SDW order and show a preference for CDW order. In addition, we found that $d$-wave pairing and spin correlation exhibit consistent behavior in the intermediate coupling regime ($U$ = 3, 4). Both remain relatively stable under attractive $V$ side but show a pronounced decline as $V$ becomes repulsive ($V > 0$).
This indicates that the effect of $V$ on the spin correlation is similar to that of the $d$-wave pairing, in some extent proves the validness of the antiferromagnetic spin fluctuations as the pairing glue of superconductivity \cite{Dahm2009-wg,Dong2022-xh}.

\begin{figure}[!]
    \centering 
    \includegraphics[width=1\linewidth]{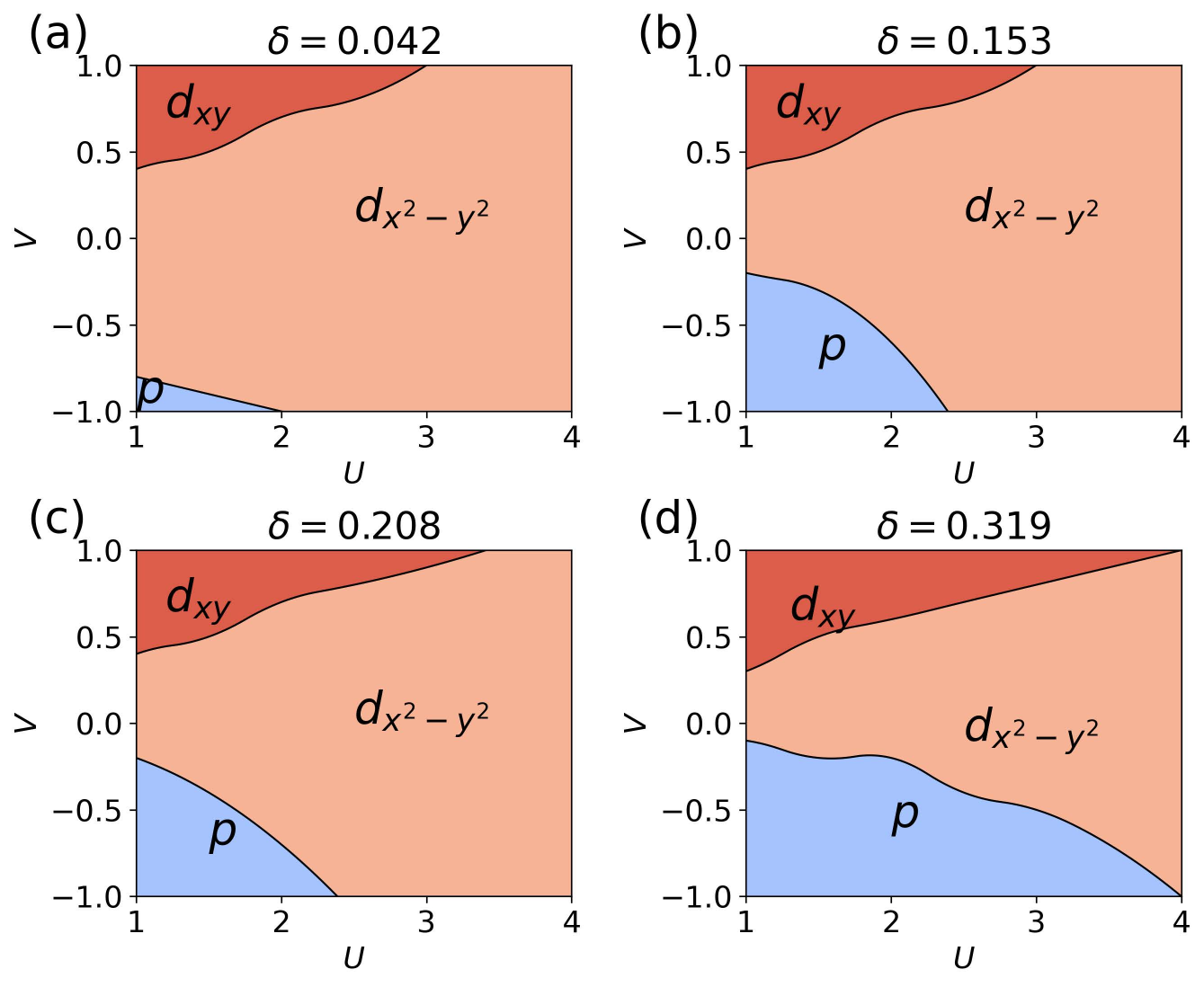}
    \caption{(Color online) $U-V$ phase diagram of different pairing modes under different doping. Schematic zero-temperature phase diagram at (a) $\delta$ = 0.042, (b) $\delta$ = 0.153, (c) $\delta$ = 0.208, (d) $\delta$ = 0.319 on 12 $\times$ 12 lattice, there are $d$-wave, $d_{xy}$-wave and exotic $p$-wave triplet pairing phases.  }
    \label{fig4}
\end{figure}

We now shift our focus to variations in doping and systematically examine an underdoped case ($\delta$ = 0.042), an optimally doped case ($\delta$ = 0.153), an overdoped case ($\delta$ = 0.208), and a heavily overdoped case ($\delta$ = 0.319) to explore the corresponding phase diagrams across different doping levels.
Overall, the phase diagrams exhibit similar characteristics under different electron doping conditions.
Specifically, when $V$ is present, a $p$-wave triplet pairing phase emerges on the attractive $V$ side, and a $d_{xy}$-wave pairing phase appears on the repulsive $V$ side. 
As $V$ increases, the regions dominated by $p$-wave and $d_{xy}$-wave pairing all expand in the phase diagram. 
As doping increases, the dominant pairing regions of $p$-wave and $d_{xy}$-wave further expand, both of which suppress $d$-wave pairing, especially in the heavily overdoped regime, as shown in Fig.\ \ref{fig4}. 
In underdoped and optimally doped systems, strong electron correlations and antiferromagnetic fluctuations typically stabilize $d$-wave pairing, however, as doping progresses into the overdoped regime, these correlations diminish. This allows other pairing channels, such as $p$-wave and $d_{xy}$-wave, to become increasingly significant.
This suggests a rich interplay of electronic interactions ($U$ and $V$), pairing symmetry, and doping-induced changes in the underlying physical properties of the system.



\begin{figure}[!]
    \centering 
    \includegraphics[width=1\linewidth]{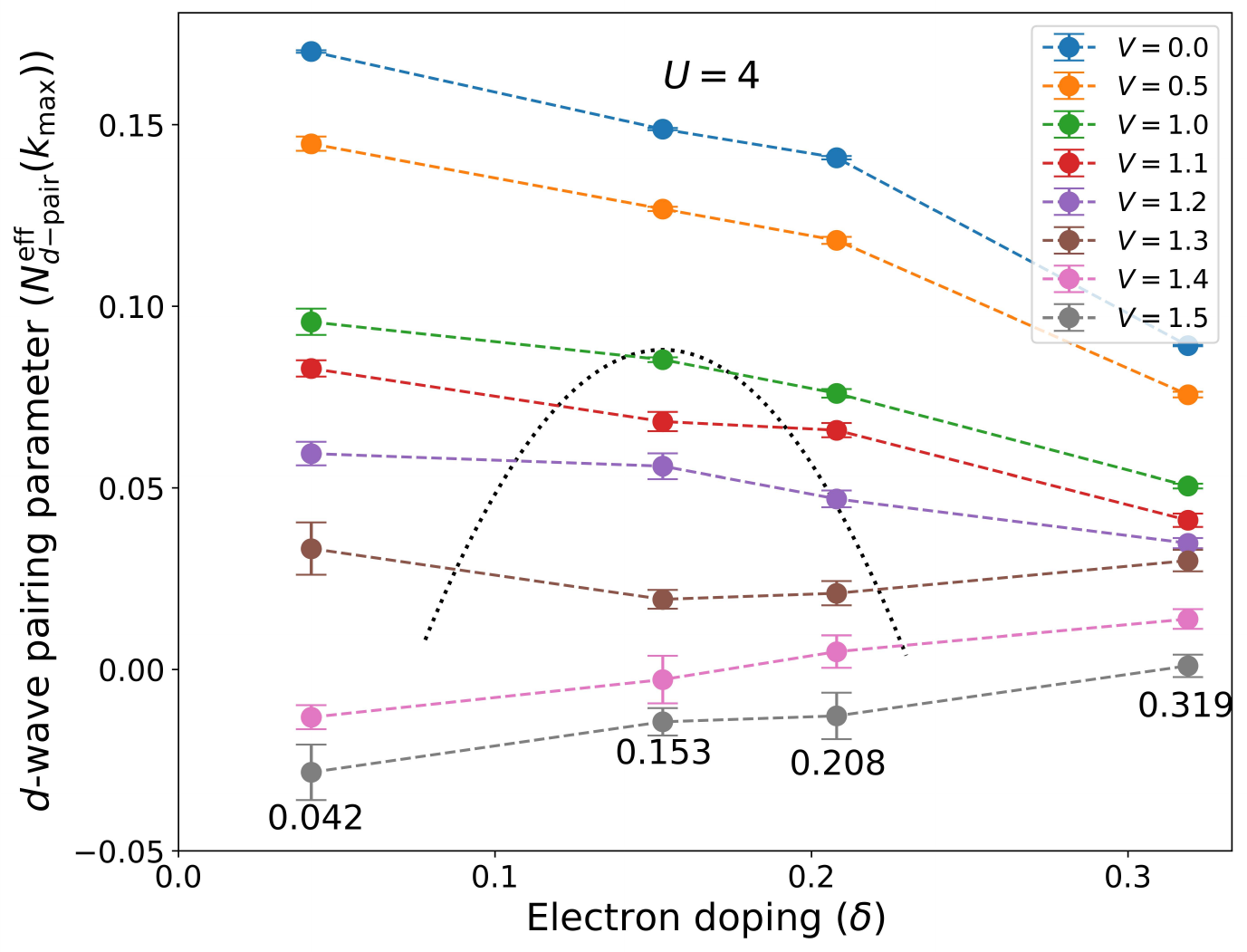}
    \caption{(Color online) The $d$-wave pairing parameter as a function of doping $\delta$ with different NN repulsion $V$. The variation of microscopic $V$ against doping is suggested by the change of magnitude in $d$-wave pairing parameter, when comparing to the trend of the dome-like shape represented by the black dotted line resembling $T_c$ domes in the typical phase diagram of cuprates.}
    \label{fig5}
\end{figure}

We analyzed the $d$-wave pairing parameter under varying doping and repulsion $V$, as depicted in Fig.\ \ref{fig5}. \zhang{Throughout this analysis, the on-site repulsion was fixed at $U = 4$, a representative value known to qualitatively capture the behavior of cuprate superconductors, which can balance physical relevance and computational feasibility \cite{Chang2010-za,Zhang2022-ej,Cao2024-hj}.}
On the whole, larger values of $V$ correspond to smaller $d$-wave pairing parameters, indicating a suppression of $d$-wave pairing strength with increasing $V$. The variation in $V$ is particularly pronounced for $1.0 \leq V \leq 1.5$. Specifically for $V < 1.3$, the $d$-wave pairing parameter gradually decreases as doping increases. At $V = 1.4$, the $d$-wave pairing parameter approaches zero, indicating significant suppression of $d$-wave pairing. Especially for $V = 1.5$, it becomes negative, suggesting that the system no longer exhibits $d$-wave pairing under this condition. In Fig.\ \ref{fig5}, we draw a black dotted dome-like shaped curve to better analyze the effects of doping $\delta$ and $V$ on the $d$-wave pairing parameters. By adjusting $V$ for different doping levels, it is possible to fit this dome-shaped curve, demonstrating the critical role of $V$ in capturing the superconducting behavior of cuprates. 
\zhang{This may be due to the variation of electron doping concentration, which leads to a change in the number of electrons in the copper oxygen surface, and the electron NN repulsion $V$ will also change with doping.} 
\zhang{Here, assuming that the suppression of the electron-doped SC region arises solely from the NN repulsion V, the ratio V/U may lie in the range of $\sim 1/4$ to 1/3, depending on the doping level. However, in practice, various other factors also contribute, such as the NNN hopping $t^{\prime}$.}
We hope that our results can provide some guidance for the exploration of the controversial electron-doped phase diagram \cite{Armitage2010-kn,Song2017-bh,Sobota2021-du}.

\section{SUMMARY}
\label{SUMMARY}

In conclusion, we employed CPQMC calculations of the 2D extended Hubbard model to investigate the impact of NN electron interactions ($V$) on the pairing phases and density wave states. At $\delta$ = 0.153, our study uncovered a quantum phase diagram encompassing $d$-wave, $d_{xy}$-wave and $p$-wave triplet pairing phases. Notably, the $p$-wave pairing phase is entirely induced and expanded by attractive $V$, while the $d$-wave pairing slowly decreases continuously as $V$ increases. However, on the repulsive $V$ side, the $d_{xy}$-wave pairing regime progressively enlarges, driven by the enhancement of $d_{xy}$-wave pairing and the suppression of $d$-wave pairing by repulsion $V$. 
Besides the pairing phase, we also find the CDW and SDW have influenced by $V$ in the particle-hole channel. Specifically, the NN electron attractive $V$ has no significant effect on SDW and CDW, but repulsive $V$ significantly enhanced CDW and suppressed SDW. 
Additionally, we found that spin correlations and $d$-wave pairing exhibit consistent behavior in the intermediate coupling regime. This consistency reinforces the validity of spin fluctuations as the dominant microscopic mechanism driving the unconventional $d$-wave SC phase.

Moreover, as doping increases, the dominant pairing region of $p$-wave and $d_{xy}$-wave also expands, further suppressing the presence of $d$-wave.
More importantly, our work suggests the repulsive $V$ is unfavorable for $d$-wave pairing and enhance the $d_{xy}$-wave pairing, which indicates that the $t-U-V$ Hubbard model with NN electron repulsive $V$ can serve as the minimal model to capture the essential physics of the electron-doped cuprates, potentially explaining the shrinkage of $d$-wave SC region and the decrease of $T_c$.

\section*{Acknowledgments}

This work is supported by the National Natural Science Foundation of China~(Grant No. 12204130), Shenzhen Start-Up Research Funds~(Grant No. HA11409065), HITSZ Start-Up Funds~(Grant No. X2022000), Shenzhen Key Laboratory of Advanced Functional CarbonMaterials Research and Comprehensive Application (Grant No. ZDSYS20220527171407017). T.Y. acknowledges supports from Natural Science Foundation of Heilongjiang Province~(No.~YQ2023A004).

\section*{DATA AVAILABILITY}
The data are available from the authors upon reasonable request.

\bibliography{ref}

\begin{thebibliography}{59}%
\makeatletter
\providecommand \@ifxundefined [1]{%
 \@ifx{#1\undefined}
}%
\providecommand \@ifnum [1]{%
 \ifnum #1\expandafter \@firstoftwo
 \else \expandafter \@secondoftwo
 \fi
}%
\providecommand \@ifx [1]{%
 \ifx #1\expandafter \@firstoftwo
 \else \expandafter \@secondoftwo
 \fi
}%
\providecommand \natexlab [1]{#1}%
\providecommand \enquote  [1]{``#1''}%
\providecommand \bibnamefont  [1]{#1}%
\providecommand \bibfnamefont [1]{#1}%
\providecommand \citenamefont [1]{#1}%
\providecommand \href@noop [0]{\@secondoftwo}%
\providecommand \href [0]{\begingroup \@sanitize@url \@href}%
\providecommand \@href[1]{\@@startlink{#1}\@@href}%
\providecommand \@@href[1]{\endgroup#1\@@endlink}%
\providecommand \@sanitize@url [0]{\catcode `\\12\catcode `\$12\catcode `\&12\catcode `\#12\catcode `\^12\catcode `\_12\catcode `\%12\relax}%
\providecommand \@@startlink[1]{}%
\providecommand \@@endlink[0]{}%
\providecommand \url  [0]{\begingroup\@sanitize@url \@url }%
\providecommand \@url [1]{\endgroup\@href {#1}{\urlprefix }}%
\providecommand \urlprefix  [0]{URL }%
\providecommand \Eprint [0]{\href }%
\providecommand \doibase [0]{https://doi.org/}%
\providecommand \selectlanguage [0]{\@gobble}%
\providecommand \bibinfo  [0]{\@secondoftwo}%
\providecommand \bibfield  [0]{\@secondoftwo}%
\providecommand \translation [1]{[#1]}%
\providecommand \BibitemOpen [0]{}%
\providecommand \bibitemStop [0]{}%
\providecommand \bibitemNoStop [0]{.\EOS\space}%
\providecommand \EOS [0]{\spacefactor3000\relax}%
\providecommand \BibitemShut  [1]{\csname bibitem#1\endcsname}%
\let\auto@bib@innerbib\@empty
\bibitem [{\citenamefont {Bednorz}\ and\ \citenamefont {M{\"u}ller}(1986)}]{Bednorz1986}%
  \BibitemOpen
  \bibfield  {author} {\bibinfo {author} {\bibfnamefont {J.~G.}\ \bibnamefont {Bednorz}}\ and\ \bibinfo {author} {\bibfnamefont {K.~A.}\ \bibnamefont {M{\"u}ller}},\ }\bibfield  {title} {\bibinfo {title} {Possible high ${T}_{c}$ superconductivity in the {Ba}$-${La}$-${Cu}$-${O} system},\ }\href {https://doi.org/10.1007/BF01303701} {\bibfield  {journal} {\bibinfo  {journal} {Z Physik B}\ }\textbf {\bibinfo {volume} {64}},\ \bibinfo {pages} {189} (\bibinfo {year} {1986})}\BibitemShut {NoStop}%
\bibitem [{\citenamefont {Tsuei}\ and\ \citenamefont {Kirtley}(2000)}]{Tsuei2000-fb}%
  \BibitemOpen
  \bibfield  {author} {\bibinfo {author} {\bibfnamefont {C.~C.}\ \bibnamefont {Tsuei}}\ and\ \bibinfo {author} {\bibfnamefont {J.~R.}\ \bibnamefont {Kirtley}},\ }\bibfield  {title} {\bibinfo {title} {{Pairing symmetry in cuprate superconductors}},\ }\href {https://doi.org/10.1103/RevModPhys.72.969} {\bibfield  {journal} {\bibinfo  {journal} {Reviews of modern physics}\ }\textbf {\bibinfo {volume} {72}},\ \bibinfo {pages} {969} (\bibinfo {year} {2000})}\BibitemShut {NoStop}%
\bibitem [{\citenamefont {Imada}\ \emph {et~al.}(1998)\citenamefont {Imada}, \citenamefont {Fujimori},\ and\ \citenamefont {Tokura}}]{Imada1998-fh}%
  \BibitemOpen
  \bibfield  {author} {\bibinfo {author} {\bibfnamefont {M.}~\bibnamefont {Imada}}, \bibinfo {author} {\bibfnamefont {A.}~\bibnamefont {Fujimori}},\ and\ \bibinfo {author} {\bibfnamefont {Y.}~\bibnamefont {Tokura}},\ }\bibfield  {title} {\bibinfo {title} {{Metal-insulator transitions}},\ }\href {https://doi.org/10.1103/revmodphys.70.1039} {\bibfield  {journal} {\bibinfo  {journal} {Reviews of modern physics}\ }\textbf {\bibinfo {volume} {70}},\ \bibinfo {pages} {1039} (\bibinfo {year} {1998})}\BibitemShut {NoStop}%
\bibitem [{\citenamefont {Orenstein}\ and\ \citenamefont {Millis}(2000)}]{Orenstein2000-am}%
  \BibitemOpen
  \bibfield  {author} {\bibinfo {author} {\bibfnamefont {J.}~\bibnamefont {Orenstein}}\ and\ \bibinfo {author} {\bibfnamefont {A.~J.}\ \bibnamefont {Millis}},\ }\bibfield  {title} {\bibinfo {title} {{Advances in the physics of high-temperature superconductivity}},\ }\href {https://doi.org/10.1126/science.288.5465.468} {\bibfield  {journal} {\bibinfo  {journal} {Science (New York, N.Y.)}\ }\textbf {\bibinfo {volume} {288}},\ \bibinfo {pages} {468} (\bibinfo {year} {2000})}\BibitemShut {NoStop}%
\bibitem [{\citenamefont {Lee}\ \emph {et~al.}(2006)\citenamefont {Lee}, \citenamefont {Nagaosa},\ and\ \citenamefont {Wen}}]{Lee2006-qq}%
  \BibitemOpen
  \bibfield  {author} {\bibinfo {author} {\bibfnamefont {P.~A.}\ \bibnamefont {Lee}}, \bibinfo {author} {\bibfnamefont {N.}~\bibnamefont {Nagaosa}},\ and\ \bibinfo {author} {\bibfnamefont {X.-G.}\ \bibnamefont {Wen}},\ }\bibfield  {title} {\bibinfo {title} {{Doping a Mott insulator: Physics of high-temperature superconductivity}},\ }\href {https://doi.org/10.1103/RevModPhys.78.17} {\bibfield  {journal} {\bibinfo  {journal} {Reviews of modern physics}\ }\textbf {\bibinfo {volume} {78}},\ \bibinfo {pages} {17} (\bibinfo {year} {2006})}\BibitemShut {NoStop}%
\bibitem [{\citenamefont {Armitage}\ \emph {et~al.}(2010)\citenamefont {Armitage}, \citenamefont {Fournier},\ and\ \citenamefont {Greene}}]{Armitage2010-kn}%
  \BibitemOpen
  \bibfield  {author} {\bibinfo {author} {\bibfnamefont {N.~P.}\ \bibnamefont {Armitage}}, \bibinfo {author} {\bibfnamefont {P.}~\bibnamefont {Fournier}},\ and\ \bibinfo {author} {\bibfnamefont {R.~L.}\ \bibnamefont {Greene}},\ }\bibfield  {title} {\bibinfo {title} {{Progress and perspectives on electron-doped cuprates}},\ }\href {https://doi.org/10.1103/revmodphys.82.2421} {\bibfield  {journal} {\bibinfo  {journal} {Reviews of modern physics}\ }\textbf {\bibinfo {volume} {82}},\ \bibinfo {pages} {2421} (\bibinfo {year} {2010})}\BibitemShut {NoStop}%
\bibitem [{\citenamefont {Fradkin}\ \emph {et~al.}(2015)\citenamefont {Fradkin}, \citenamefont {Kivelson},\ and\ \citenamefont {Tranquada}}]{Fradkin2015-xz}%
  \BibitemOpen
  \bibfield  {author} {\bibinfo {author} {\bibfnamefont {E.}~\bibnamefont {Fradkin}}, \bibinfo {author} {\bibfnamefont {S.~A.}\ \bibnamefont {Kivelson}},\ and\ \bibinfo {author} {\bibfnamefont {J.~M.}\ \bibnamefont {Tranquada}},\ }\bibfield  {title} {\bibinfo {title} {{Colloquium: Theory of intertwined orders in high temperature superconductors}},\ }\href {https://doi.org/10.1103/RevModPhys.87.457} {\bibfield  {journal} {\bibinfo  {journal} {Reviews of modern physics}\ }\textbf {\bibinfo {volume} {87}},\ \bibinfo {pages} {457} (\bibinfo {year} {2015})}\BibitemShut {NoStop}%
\bibitem [{\citenamefont {Zhou}\ \emph {et~al.}(2021)\citenamefont {Zhou}, \citenamefont {Lee}, \citenamefont {Imada}, \citenamefont {Trivedi}, \citenamefont {Phillips}, \citenamefont {Kee}, \citenamefont {T{\"{o}}rm{\"{a}}},\ and\ \citenamefont {Eremets}}]{Zhou2021-xg}%
  \BibitemOpen
  \bibfield  {author} {\bibinfo {author} {\bibfnamefont {X.}~\bibnamefont {Zhou}}, \bibinfo {author} {\bibfnamefont {W.-S.}\ \bibnamefont {Lee}}, \bibinfo {author} {\bibfnamefont {M.}~\bibnamefont {Imada}}, \bibinfo {author} {\bibfnamefont {N.}~\bibnamefont {Trivedi}}, \bibinfo {author} {\bibfnamefont {P.}~\bibnamefont {Phillips}}, \bibinfo {author} {\bibfnamefont {H.-Y.}\ \bibnamefont {Kee}}, \bibinfo {author} {\bibfnamefont {P.}~\bibnamefont {T{\"{o}}rm{\"{a}}}},\ and\ \bibinfo {author} {\bibfnamefont {M.}~\bibnamefont {Eremets}},\ }\bibfield  {title} {\bibinfo {title} {{High-temperature superconductivity}},\ }\href {https://doi.org/10.1038/s42254-021-00324-3} {\bibfield  {journal} {\bibinfo  {journal} {Nature reviews. Physics}\ }\textbf {\bibinfo {volume} {3}},\ \bibinfo {pages} {462} (\bibinfo {year} {2021})}\BibitemShut {NoStop}%
\bibitem [{\citenamefont {Keimer}\ \emph {et~al.}(2015)\citenamefont {Keimer}, \citenamefont {Kivelson}, \citenamefont {Norman}, \citenamefont {Uchida},\ and\ \citenamefont {Zaanen}}]{Keimer2015-jp}%
  \BibitemOpen
  \bibfield  {author} {\bibinfo {author} {\bibfnamefont {B.}~\bibnamefont {Keimer}}, \bibinfo {author} {\bibfnamefont {S.~A.}\ \bibnamefont {Kivelson}}, \bibinfo {author} {\bibfnamefont {M.~R.}\ \bibnamefont {Norman}}, \bibinfo {author} {\bibfnamefont {S.}~\bibnamefont {Uchida}},\ and\ \bibinfo {author} {\bibfnamefont {J.}~\bibnamefont {Zaanen}},\ }\bibfield  {title} {\bibinfo {title} {{From quantum matter to high-temperature superconductivity in copper oxides}},\ }\href {https://doi.org/10.1038/nature14165} {\bibfield  {journal} {\bibinfo  {journal} {Nature}\ }\textbf {\bibinfo {volume} {518}},\ \bibinfo {pages} {179} (\bibinfo {year} {2015})}\BibitemShut {NoStop}%
\bibitem [{\citenamefont {Zhang}\ and\ \citenamefont {Rice}(1988)}]{Zhang1988-dw}%
  \BibitemOpen
  \bibfield  {author} {\bibinfo {author} {\bibfnamefont {F.~C.}\ \bibnamefont {Zhang}}\ and\ \bibinfo {author} {\bibfnamefont {T.~M.}\ \bibnamefont {Rice}},\ }\bibfield  {title} {\bibinfo {title} {{Effective Hamiltonian for the superconducting Cu oxides}},\ }\href {https://doi.org/10.1103/physrevb.37.3759} {\bibfield  {journal} {\bibinfo  {journal} {Physical review. B, Condensed matter}\ }\textbf {\bibinfo {volume} {37}},\ \bibinfo {pages} {3759} (\bibinfo {year} {1988})}\BibitemShut {NoStop}%
\bibitem [{\citenamefont {Anderson}\ \emph {et~al.}(2004)\citenamefont {Anderson}, \citenamefont {Lee}, \citenamefont {Randeria}, \citenamefont {Rice}, \citenamefont {Trivedi},\ and\ \citenamefont {Zhang}}]{Anderson2004-gq}%
  \BibitemOpen
  \bibfield  {author} {\bibinfo {author} {\bibfnamefont {P.~W.}\ \bibnamefont {Anderson}}, \bibinfo {author} {\bibfnamefont {P.~A.}\ \bibnamefont {Lee}}, \bibinfo {author} {\bibfnamefont {M.}~\bibnamefont {Randeria}}, \bibinfo {author} {\bibfnamefont {T.~M.}\ \bibnamefont {Rice}}, \bibinfo {author} {\bibfnamefont {N.}~\bibnamefont {Trivedi}},\ and\ \bibinfo {author} {\bibfnamefont {F.~C.}\ \bibnamefont {Zhang}},\ }\bibfield  {title} {\bibinfo {title} {{The physics behind high-temperature superconducting cuprates: the plain vanilla version of RVB}},\ }\href {https://doi.org/10.1088/0953-8984/16/24/r02} {\bibfield  {journal} {\bibinfo  {journal} {Journal of physics. Condensed matter: an Institute of Physics journal}\ }\textbf {\bibinfo {volume} {16}},\ \bibinfo {pages} {R755} (\bibinfo {year} {2004})}\BibitemShut {NoStop}%
\bibitem [{\citenamefont {Jiang}\ \emph {et~al.}(2021)\citenamefont {Jiang}, \citenamefont {Scalapino},\ and\ \citenamefont {White}}]{Jiang2021-hj}%
  \BibitemOpen
  \bibfield  {author} {\bibinfo {author} {\bibfnamefont {S.}~\bibnamefont {Jiang}}, \bibinfo {author} {\bibfnamefont {D.~J.}\ \bibnamefont {Scalapino}},\ and\ \bibinfo {author} {\bibfnamefont {S.~R.}\ \bibnamefont {White}},\ }\bibfield  {title} {\bibinfo {title} {{Ground-state phase diagram of the \textit{t-t} ' \textit{-J} model}},\ }\bibfield  {journal} {\bibinfo  {journal} {Proceedings of the National Academy of Sciences of the United States of America}\ }\textbf {\bibinfo {volume} {118}},\ \href {https://doi.org/10.1073/pnas.2109978118} {10.1073/pnas.2109978118} (\bibinfo {year} {2021})\BibitemShut {NoStop}%
\bibitem [{\citenamefont {Qin}\ \emph {et~al.}(2022)\citenamefont {Qin}, \citenamefont {Sch{\"{a}}fer}, \citenamefont {Andergassen}, \citenamefont {Corboz},\ and\ \citenamefont {Gull}}]{Qin2022-ad}%
  \BibitemOpen
  \bibfield  {author} {\bibinfo {author} {\bibfnamefont {M.}~\bibnamefont {Qin}}, \bibinfo {author} {\bibfnamefont {T.}~\bibnamefont {Sch{\"{a}}fer}}, \bibinfo {author} {\bibfnamefont {S.}~\bibnamefont {Andergassen}}, \bibinfo {author} {\bibfnamefont {P.}~\bibnamefont {Corboz}},\ and\ \bibinfo {author} {\bibfnamefont {E.}~\bibnamefont {Gull}},\ }\bibfield  {title} {\bibinfo {title} {{The Hubbard model: A computational perspective}},\ }\href {https://doi.org/10.1146/annurev-conmatphys-090921-033948} {\bibfield  {journal} {\bibinfo  {journal} {Annual review of condensed matter physics}\ }\textbf {\bibinfo {volume} {13}},\ \bibinfo {pages} {275} (\bibinfo {year} {2022})}\BibitemShut {NoStop}%
\bibitem [{\citenamefont {Mai}\ \emph {et~al.}(2021)\citenamefont {Mai}, \citenamefont {Balduzzi}, \citenamefont {Johnston},\ and\ \citenamefont {Maier}}]{Mai2021-wv}%
  \BibitemOpen
  \bibfield  {author} {\bibinfo {author} {\bibfnamefont {P.}~\bibnamefont {Mai}}, \bibinfo {author} {\bibfnamefont {G.}~\bibnamefont {Balduzzi}}, \bibinfo {author} {\bibfnamefont {S.}~\bibnamefont {Johnston}},\ and\ \bibinfo {author} {\bibfnamefont {T.~A.}\ \bibnamefont {Maier}},\ }\bibfield  {title} {\bibinfo {title} {{Orbital structure of the effective pairing interaction in the high-temperature superconducting cuprates}},\ }\href {https://doi.org/10.1038/s41535-021-00326-5} {\bibfield  {journal} {\bibinfo  {journal} {npj Quantum Materials}\ }\textbf {\bibinfo {volume} {6}},\ \bibinfo {pages} {1} (\bibinfo {year} {2021})}\BibitemShut {NoStop}%
\bibitem [{\citenamefont {Scalapino}(2012)}]{Scalapino2012-jr}%
  \BibitemOpen
  \bibfield  {author} {\bibinfo {author} {\bibfnamefont {D.~J.}\ \bibnamefont {Scalapino}},\ }\bibfield  {title} {\bibinfo {title} {{A common thread: The pairing interaction for unconventional superconductors}},\ }\href {https://doi.org/10.1103/RevModPhys.84.1383} {\bibfield  {journal} {\bibinfo  {journal} {Reviews of modern physics}\ }\textbf {\bibinfo {volume} {84}},\ \bibinfo {pages} {1383} (\bibinfo {year} {2012})}\BibitemShut {NoStop}%
\bibitem [{\citenamefont {Sobota}\ \emph {et~al.}(2021)\citenamefont {Sobota}, \citenamefont {He},\ and\ \citenamefont {Shen}}]{Sobota2021-du}%
  \BibitemOpen
  \bibfield  {author} {\bibinfo {author} {\bibfnamefont {J.~A.}\ \bibnamefont {Sobota}}, \bibinfo {author} {\bibfnamefont {Y.}~\bibnamefont {He}},\ and\ \bibinfo {author} {\bibfnamefont {Z.-X.}\ \bibnamefont {Shen}},\ }\bibfield  {title} {\bibinfo {title} {Angle-resolved photoemission studies of quantum materials},\ }\href {https://doi.org/10.1103/RevModPhys.93.025006} {\bibfield  {journal} {\bibinfo  {journal} {Rev. Mod. Phys.}\ }\textbf {\bibinfo {volume} {93}},\ \bibinfo {pages} {025006} (\bibinfo {year} {2021})}\BibitemShut {NoStop}%
\bibitem [{\citenamefont {Jiang}\ and\ \citenamefont {Devereaux}(2019)}]{Jiang2019-km}%
  \BibitemOpen
  \bibfield  {author} {\bibinfo {author} {\bibfnamefont {H.-C.}\ \bibnamefont {Jiang}}\ and\ \bibinfo {author} {\bibfnamefont {T.~P.}\ \bibnamefont {Devereaux}},\ }\bibfield  {title} {\bibinfo {title} {{Superconductivity in the doped Hubbard model and its interplay with next-nearest hopping t'}},\ }\href {https://doi.org/10.1126/science.aal5304} {\bibfield  {journal} {\bibinfo  {journal} {Science}\ }\textbf {\bibinfo {volume} {365}},\ \bibinfo {pages} {1424} (\bibinfo {year} {2019})}\BibitemShut {NoStop}%
\bibitem [{\citenamefont {Qin}\ \emph {et~al.}(2020)\citenamefont {Qin}, \citenamefont {Chung}, \citenamefont {Shi}, \citenamefont {Vitali}, \citenamefont {Hubig}, \citenamefont {Schollw\"ock}, \citenamefont {White},\ and\ \citenamefont {Zhang}}]{qin2020absence}%
  \BibitemOpen
  \bibfield  {author} {\bibinfo {author} {\bibfnamefont {M.}~\bibnamefont {Qin}}, \bibinfo {author} {\bibfnamefont {C.-M.}\ \bibnamefont {Chung}}, \bibinfo {author} {\bibfnamefont {H.}~\bibnamefont {Shi}}, \bibinfo {author} {\bibfnamefont {E.}~\bibnamefont {Vitali}}, \bibinfo {author} {\bibfnamefont {C.}~\bibnamefont {Hubig}}, \bibinfo {author} {\bibfnamefont {U.}~\bibnamefont {Schollw\"ock}}, \bibinfo {author} {\bibfnamefont {S.~R.}\ \bibnamefont {White}},\ and\ \bibinfo {author} {\bibfnamefont {S.}~\bibnamefont {Zhang}} (\bibinfo {collaboration} {Simons Collaboration on the Many-Electron Problem}),\ }\bibfield  {title} {\bibinfo {title} {Absence of superconductivity in the pure two-dimensional hubbard model},\ }\href {https://doi.org/10.1103/PhysRevX.10.031016} {\bibfield  {journal} {\bibinfo  {journal} {Phys. Rev. X}\ }\textbf {\bibinfo {volume} {10}},\ \bibinfo {pages} {031016} (\bibinfo {year} {2020})}\BibitemShut {NoStop}%
\bibitem [{\citenamefont {Chen}\ \emph {et~al.}(2023)\citenamefont {Chen}, \citenamefont {Haldane},\ and\ \citenamefont {Sheng}}]{Chen2023-rk}%
  \BibitemOpen
  \bibfield  {author} {\bibinfo {author} {\bibfnamefont {F.}~\bibnamefont {Chen}}, \bibinfo {author} {\bibfnamefont {F.~D.~M.}\ \bibnamefont {Haldane}},\ and\ \bibinfo {author} {\bibfnamefont {D.~N.}\ \bibnamefont {Sheng}},\ }\bibfield  {title} {\bibinfo {title} {{D-wave and pair-density-wave superconductivity in the square-lattice t-J model}},\ }\href@noop {} {\bibfield  {journal} {\bibinfo  {journal} {arXiv [cond-mat.supr-con]}\ } (\bibinfo {year} {2023})},\ \Eprint {https://arxiv.org/abs/2311.15092} {arXiv:2311.15092 [cond-mat.supr-con]} \BibitemShut {NoStop}%
\bibitem [{\citenamefont {Xu}\ \emph {et~al.}(2024)\citenamefont {Xu}, \citenamefont {Chung}, \citenamefont {Qin}, \citenamefont {Schollwöck}, \citenamefont {White},\ and\ \citenamefont {Zhang}}]{Xu2024-wt}%
  \BibitemOpen
  \bibfield  {author} {\bibinfo {author} {\bibfnamefont {H.}~\bibnamefont {Xu}}, \bibinfo {author} {\bibfnamefont {C.-M.}\ \bibnamefont {Chung}}, \bibinfo {author} {\bibfnamefont {M.}~\bibnamefont {Qin}}, \bibinfo {author} {\bibfnamefont {U.}~\bibnamefont {Schollwöck}}, \bibinfo {author} {\bibfnamefont {S.~R.}\ \bibnamefont {White}},\ and\ \bibinfo {author} {\bibfnamefont {S.}~\bibnamefont {Zhang}},\ }\bibfield  {title} {\bibinfo {title} {Coexistence of superconductivity with partially filled stripes in the hubbard model},\ }\href {https://doi.org/10.1126/science.adh7691} {\bibfield  {journal} {\bibinfo  {journal} {Science}\ }\textbf {\bibinfo {volume} {384}},\ \bibinfo {pages} {eadh7691} (\bibinfo {year} {2024})}\BibitemShut {NoStop}%
\bibitem [{\citenamefont {Lu}\ \emph {et~al.}(2024{\natexlab{a}})\citenamefont {Lu}, \citenamefont {Chen}, \citenamefont {Zhu}, \citenamefont {Sheng},\ and\ \citenamefont {Gong}}]{Lu2024-ah}%
  \BibitemOpen
  \bibfield  {author} {\bibinfo {author} {\bibfnamefont {X.}~\bibnamefont {Lu}}, \bibinfo {author} {\bibfnamefont {F.}~\bibnamefont {Chen}}, \bibinfo {author} {\bibfnamefont {W.}~\bibnamefont {Zhu}}, \bibinfo {author} {\bibfnamefont {D.~N.}\ \bibnamefont {Sheng}},\ and\ \bibinfo {author} {\bibfnamefont {S.-S.}\ \bibnamefont {Gong}},\ }\bibfield  {title} {\bibinfo {title} {{Emergent superconductivity and competing charge orders in hole-doped square-lattice t-J model}},\ }\href {https://doi.org/10.1103/PhysRevLett.132.066002} {\bibfield  {journal} {\bibinfo  {journal} {Physical review letters}\ }\textbf {\bibinfo {volume} {132}},\ \bibinfo {pages} {066002} (\bibinfo {year} {2024}{\natexlab{a}})}\BibitemShut {NoStop}%
\bibitem [{\citenamefont {Motoyama}\ \emph {et~al.}(2007)\citenamefont {Motoyama}, \citenamefont {Yu}, \citenamefont {Vishik}, \citenamefont {Vajk}, \citenamefont {Mang},\ and\ \citenamefont {Greven}}]{Motoyama2007-th}%
  \BibitemOpen
  \bibfield  {author} {\bibinfo {author} {\bibfnamefont {E.~M.}\ \bibnamefont {Motoyama}}, \bibinfo {author} {\bibfnamefont {G.}~\bibnamefont {Yu}}, \bibinfo {author} {\bibfnamefont {I.~M.}\ \bibnamefont {Vishik}}, \bibinfo {author} {\bibfnamefont {O.~P.}\ \bibnamefont {Vajk}}, \bibinfo {author} {\bibfnamefont {P.~K.}\ \bibnamefont {Mang}},\ and\ \bibinfo {author} {\bibfnamefont {M.}~\bibnamefont {Greven}},\ }\bibfield  {title} {\bibinfo {title} {{Spin correlations in the electron-doped high-transition-temperature superconductor Nd$_{2-x}$Ce${_x}$CuO$_{4\pm\delta}$}},\ }\href {https://doi.org/10.1038/nature05437} {\bibfield  {journal} {\bibinfo  {journal} {Nature}\ }\textbf {\bibinfo {volume} {445}},\ \bibinfo {pages} {186} (\bibinfo {year} {2007})}\BibitemShut {NoStop}%
\bibitem [{\citenamefont {Chen}\ \emph {et~al.}(2024)\citenamefont {Chen}, \citenamefont {Qiao}, \citenamefont {Zhang},\ and\ \citenamefont {Zhu}}]{Chen2024-rr}%
  \BibitemOpen
  \bibfield  {author} {\bibinfo {author} {\bibfnamefont {Q.}~\bibnamefont {Chen}}, \bibinfo {author} {\bibfnamefont {L.}~\bibnamefont {Qiao}}, \bibinfo {author} {\bibfnamefont {F.}~\bibnamefont {Zhang}},\ and\ \bibinfo {author} {\bibfnamefont {Z.}~\bibnamefont {Zhu}},\ }\bibfield  {title} {\bibinfo {title} {Phase diagram of the square-lattice $t-j-v$ model for electron-doped cuprates},\ }\href {https://doi.org/10.1103/PhysRevB.110.045134} {\bibfield  {journal} {\bibinfo  {journal} {Phys. Rev. B}\ }\textbf {\bibinfo {volume} {110}},\ \bibinfo {pages} {045134} (\bibinfo {year} {2024})}\BibitemShut {NoStop}%
\bibitem [{\citenamefont {Uefuji}\ \emph {et~al.}(2001)\citenamefont {Uefuji}, \citenamefont {Kubo}, \citenamefont {Yamada}, \citenamefont {Fujita}, \citenamefont {Kurahashi}, \citenamefont {Watanabe},\ and\ \citenamefont {Nagamine}}]{Uefuji2001-vc}%
  \BibitemOpen
  \bibfield  {author} {\bibinfo {author} {\bibfnamefont {T.}~\bibnamefont {Uefuji}}, \bibinfo {author} {\bibfnamefont {T.}~\bibnamefont {Kubo}}, \bibinfo {author} {\bibfnamefont {K.}~\bibnamefont {Yamada}}, \bibinfo {author} {\bibfnamefont {M.}~\bibnamefont {Fujita}}, \bibinfo {author} {\bibfnamefont {K.}~\bibnamefont {Kurahashi}}, \bibinfo {author} {\bibfnamefont {I.}~\bibnamefont {Watanabe}},\ and\ \bibinfo {author} {\bibfnamefont {K.}~\bibnamefont {Nagamine}},\ }\bibfield  {title} {\bibinfo {title} {{Coexistence of antiferromagnetic ordering and high-Tc superconductivity in electron-doped superconductor Nd2-Ce CuO4}},\ }\href {https://doi.org/10.1016/s0921-4534(01)00208-8} {\bibfield  {journal} {\bibinfo  {journal} {Physica. C, Superconductivity}\ }\textbf {\bibinfo {volume} {357-360}},\ \bibinfo {pages} {208} (\bibinfo {year} {2001})}\BibitemShut {NoStop}%
\bibitem [{\citenamefont {Comin}\ \emph {et~al.}(2014)\citenamefont {Comin}, \citenamefont {Frano}, \citenamefont {Yee}, \citenamefont {Yoshida}, \citenamefont {Eisaki}, \citenamefont {Schierle}, \citenamefont {Weschke}, \citenamefont {Sutarto}, \citenamefont {He}, \citenamefont {Soumyanarayanan}, \citenamefont {He}, \citenamefont {Le~Tacon}, \citenamefont {Elfimov}, \citenamefont {Hoffman}, \citenamefont {Sawatzky}, \citenamefont {Keimer},\ and\ \citenamefont {Damascelli}}]{Comin2014-bz}%
  \BibitemOpen
  \bibfield  {author} {\bibinfo {author} {\bibfnamefont {R.}~\bibnamefont {Comin}}, \bibinfo {author} {\bibfnamefont {A.}~\bibnamefont {Frano}}, \bibinfo {author} {\bibfnamefont {M.~M.}\ \bibnamefont {Yee}}, \bibinfo {author} {\bibfnamefont {Y.}~\bibnamefont {Yoshida}}, \bibinfo {author} {\bibfnamefont {H.}~\bibnamefont {Eisaki}}, \bibinfo {author} {\bibfnamefont {E.}~\bibnamefont {Schierle}}, \bibinfo {author} {\bibfnamefont {E.}~\bibnamefont {Weschke}}, \bibinfo {author} {\bibfnamefont {R.}~\bibnamefont {Sutarto}}, \bibinfo {author} {\bibfnamefont {F.}~\bibnamefont {He}}, \bibinfo {author} {\bibfnamefont {A.}~\bibnamefont {Soumyanarayanan}}, \bibinfo {author} {\bibfnamefont {Y.}~\bibnamefont {He}}, \bibinfo {author} {\bibfnamefont {M.}~\bibnamefont {Le~Tacon}}, \bibinfo {author} {\bibfnamefont {I.~S.}\ \bibnamefont {Elfimov}}, \bibinfo {author} {\bibfnamefont {J.~E.}\ \bibnamefont {Hoffman}}, \bibinfo {author} {\bibfnamefont {G.~A.}\ \bibnamefont {Sawatzky}}, \bibinfo {author} {\bibfnamefont
  {B.}~\bibnamefont {Keimer}},\ and\ \bibinfo {author} {\bibfnamefont {A.}~\bibnamefont {Damascelli}},\ }\bibfield  {title} {\bibinfo {title} {{Charge order driven by Fermi-arc instability in Bi$_{3}$Sr$_{2-x}$La$_{x}$CuO$_{6+\delta}$}},\ }\href {https://doi.org/10.1126/science.1242996} {\bibfield  {journal} {\bibinfo  {journal} {Science}\ }\textbf {\bibinfo {volume} {343}},\ \bibinfo {pages} {390} (\bibinfo {year} {2014})}\BibitemShut {NoStop}%
\bibitem [{\citenamefont {Moon}\ and\ \citenamefont {Sachdev}(2009)}]{Moon2009-il}%
  \BibitemOpen
  \bibfield  {author} {\bibinfo {author} {\bibfnamefont {E.~G.}\ \bibnamefont {Moon}}\ and\ \bibinfo {author} {\bibfnamefont {S.}~\bibnamefont {Sachdev}},\ }\bibfield  {title} {\bibinfo {title} {{Competition between spin density wave order and superconductivity in the underdoped cuprates}},\ }\href {https://doi.org/10.1103/PhysRevB.80.035117} {\bibfield  {journal} {\bibinfo  {journal} {Physical review. B, Condensed matter}\ }\textbf {\bibinfo {volume} {80}},\ \bibinfo {pages} {035117} (\bibinfo {year} {2009})}\BibitemShut {NoStop}%
\bibitem [{\citenamefont {Shi}\ \emph {et~al.}(2020)\citenamefont {Shi}, \citenamefont {Baity}, \citenamefont {Terzic}, \citenamefont {Sasagawa},\ and\ \citenamefont {Popović}}]{Shi2020-vh}%
  \BibitemOpen
  \bibfield  {author} {\bibinfo {author} {\bibfnamefont {Z.}~\bibnamefont {Shi}}, \bibinfo {author} {\bibfnamefont {P.~G.}\ \bibnamefont {Baity}}, \bibinfo {author} {\bibfnamefont {J.}~\bibnamefont {Terzic}}, \bibinfo {author} {\bibfnamefont {T.}~\bibnamefont {Sasagawa}},\ and\ \bibinfo {author} {\bibfnamefont {D.}~\bibnamefont {Popović}},\ }\bibfield  {title} {\bibinfo {title} {{Pair density wave at high magnetic fields in cuprates with charge and spin orders}},\ }\href {https://doi.org/10.1038/s41467-020-17138-z} {\bibfield  {journal} {\bibinfo  {journal} {Nature communications}\ }\textbf {\bibinfo {volume} {11}},\ \bibinfo {pages} {3323} (\bibinfo {year} {2020})}\BibitemShut {NoStop}%
\bibitem [{\citenamefont {Sato}\ \emph {et~al.}(2017)\citenamefont {Sato}, \citenamefont {Kasahara}, \citenamefont {Murayama}, \citenamefont {Kasahara}, \citenamefont {Moon}, \citenamefont {Nishizaki}, \citenamefont {Loew}, \citenamefont {Porras}, \citenamefont {Keimer}, \citenamefont {Shibauchi},\ and\ \citenamefont {Matsuda}}]{Sato2017-bt}%
  \BibitemOpen
  \bibfield  {author} {\bibinfo {author} {\bibfnamefont {Y.}~\bibnamefont {Sato}}, \bibinfo {author} {\bibfnamefont {S.}~\bibnamefont {Kasahara}}, \bibinfo {author} {\bibfnamefont {H.}~\bibnamefont {Murayama}}, \bibinfo {author} {\bibfnamefont {Y.}~\bibnamefont {Kasahara}}, \bibinfo {author} {\bibfnamefont {E.-G.}\ \bibnamefont {Moon}}, \bibinfo {author} {\bibfnamefont {T.}~\bibnamefont {Nishizaki}}, \bibinfo {author} {\bibfnamefont {T.}~\bibnamefont {Loew}}, \bibinfo {author} {\bibfnamefont {J.}~\bibnamefont {Porras}}, \bibinfo {author} {\bibfnamefont {B.}~\bibnamefont {Keimer}}, \bibinfo {author} {\bibfnamefont {T.}~\bibnamefont {Shibauchi}},\ and\ \bibinfo {author} {\bibfnamefont {Y.}~\bibnamefont {Matsuda}},\ }\bibfield  {title} {\bibinfo {title} {{Thermodynamic evidence for a nematic phase transition at the onset of the pseudogap in YBa$_{2}$Cu$_{3}$O$_{y}$}},\ }\href {https://doi.org/10.1038/nphys4205} {\bibfield  {journal} {\bibinfo  {journal} {Nature physics}\ }\textbf {\bibinfo {volume} {13}},\
  \bibinfo {pages} {1074} (\bibinfo {year} {2017})}\BibitemShut {NoStop}%
\bibitem [{\citenamefont {Doiron-Leyraud}\ \emph {et~al.}(2017)\citenamefont {Doiron-Leyraud}, \citenamefont {Cyr-Choini\`{e}re}, \citenamefont {Badoux}, \citenamefont {Ataei}, \citenamefont {Collignon}, \citenamefont {Gourgout}, \citenamefont {Dufour-Beaus\'{e}jour}, \citenamefont {Tafti}, \citenamefont {Lalibert\'{e}}, \citenamefont {Boulanger}, \citenamefont {Matusiak}, \citenamefont {Graf}, \citenamefont {Kim}, \citenamefont {Zhou}, \citenamefont {Momono}, \citenamefont {Kurosawa}, \citenamefont {Takagi},\ and\ \citenamefont {Taillefer}}]{Doiron-Leyraud2017-lv}%
  \BibitemOpen
  \bibfield  {author} {\bibinfo {author} {\bibfnamefont {N.}~\bibnamefont {Doiron-Leyraud}}, \bibinfo {author} {\bibfnamefont {O.}~\bibnamefont {Cyr-Choini\`{e}re}}, \bibinfo {author} {\bibfnamefont {S.}~\bibnamefont {Badoux}}, \bibinfo {author} {\bibfnamefont {A.}~\bibnamefont {Ataei}}, \bibinfo {author} {\bibfnamefont {C.}~\bibnamefont {Collignon}}, \bibinfo {author} {\bibfnamefont {A.}~\bibnamefont {Gourgout}}, \bibinfo {author} {\bibfnamefont {S.}~\bibnamefont {Dufour-Beaus\'{e}jour}}, \bibinfo {author} {\bibfnamefont {F.~F.}\ \bibnamefont {Tafti}}, \bibinfo {author} {\bibfnamefont {F.}~\bibnamefont {Lalibert\'{e}}}, \bibinfo {author} {\bibfnamefont {M.-E.}\ \bibnamefont {Boulanger}}, \bibinfo {author} {\bibfnamefont {M.}~\bibnamefont {Matusiak}}, \bibinfo {author} {\bibfnamefont {D.}~\bibnamefont {Graf}}, \bibinfo {author} {\bibfnamefont {M.}~\bibnamefont {Kim}}, \bibinfo {author} {\bibfnamefont {J.-S.}\ \bibnamefont {Zhou}}, \bibinfo {author} {\bibfnamefont {N.}~\bibnamefont {Momono}}, \bibinfo {author}
  {\bibfnamefont {T.}~\bibnamefont {Kurosawa}}, \bibinfo {author} {\bibfnamefont {H.}~\bibnamefont {Takagi}},\ and\ \bibinfo {author} {\bibfnamefont {L.}~\bibnamefont {Taillefer}},\ }\bibfield  {title} {\bibinfo {title} {{Pseudogap phase of cuprate superconductors confined by Fermi surface topology}},\ }\href {https://doi.org/10.1038/s41467-017-02122-x} {\bibfield  {journal} {\bibinfo  {journal} {Nature communications}\ }\textbf {\bibinfo {volume} {8}},\ \bibinfo {pages} {1} (\bibinfo {year} {2017})}\BibitemShut {NoStop}%
\bibitem [{\citenamefont {He}\ \emph {et~al.}(2014)\citenamefont {He}, \citenamefont {Yin}, \citenamefont {Zech}, \citenamefont {Soumyanarayanan}, \citenamefont {Yee}, \citenamefont {Williams}, \citenamefont {Boyer}, \citenamefont {Chatterjee}, \citenamefont {Wise}, \citenamefont {Zeljkovic}, \citenamefont {Kondo}, \citenamefont {Takeuchi}, \citenamefont {Ikuta}, \citenamefont {Mistark}, \citenamefont {Markiewicz}, \citenamefont {Bansil}, \citenamefont {Sachdev}, \citenamefont {Hudson},\ and\ \citenamefont {Hoffman}}]{He2014-bx}%
  \BibitemOpen
  \bibfield  {author} {\bibinfo {author} {\bibfnamefont {Y.}~\bibnamefont {He}}, \bibinfo {author} {\bibfnamefont {Y.}~\bibnamefont {Yin}}, \bibinfo {author} {\bibfnamefont {M.}~\bibnamefont {Zech}}, \bibinfo {author} {\bibfnamefont {A.}~\bibnamefont {Soumyanarayanan}}, \bibinfo {author} {\bibfnamefont {M.~M.}\ \bibnamefont {Yee}}, \bibinfo {author} {\bibfnamefont {T.}~\bibnamefont {Williams}}, \bibinfo {author} {\bibfnamefont {M.~C.}\ \bibnamefont {Boyer}}, \bibinfo {author} {\bibfnamefont {K.}~\bibnamefont {Chatterjee}}, \bibinfo {author} {\bibfnamefont {W.~D.}\ \bibnamefont {Wise}}, \bibinfo {author} {\bibfnamefont {I.}~\bibnamefont {Zeljkovic}}, \bibinfo {author} {\bibfnamefont {T.}~\bibnamefont {Kondo}}, \bibinfo {author} {\bibfnamefont {T.}~\bibnamefont {Takeuchi}}, \bibinfo {author} {\bibfnamefont {H.}~\bibnamefont {Ikuta}}, \bibinfo {author} {\bibfnamefont {P.}~\bibnamefont {Mistark}}, \bibinfo {author} {\bibfnamefont {R.~S.}\ \bibnamefont {Markiewicz}}, \bibinfo {author} {\bibfnamefont
  {A.}~\bibnamefont {Bansil}}, \bibinfo {author} {\bibfnamefont {S.}~\bibnamefont {Sachdev}}, \bibinfo {author} {\bibfnamefont {E.~W.}\ \bibnamefont {Hudson}},\ and\ \bibinfo {author} {\bibfnamefont {J.~E.}\ \bibnamefont {Hoffman}},\ }\bibfield  {title} {\bibinfo {title} {{Fermi surface and pseudogap evolution in a cuprate superconductor}},\ }\href {https://doi.org/10.1126/science.1248221} {\bibfield  {journal} {\bibinfo  {journal} {Science (New York, N.Y.)}\ }\textbf {\bibinfo {volume} {344}},\ \bibinfo {pages} {608} (\bibinfo {year} {2014})}\BibitemShut {NoStop}%
\bibitem [{\citenamefont {Himeda}\ \emph {et~al.}(2002)\citenamefont {Himeda}, \citenamefont {Kato},\ and\ \citenamefont {Ogata}}]{Himeda2002-mc}%
  \BibitemOpen
  \bibfield  {author} {\bibinfo {author} {\bibfnamefont {A.}~\bibnamefont {Himeda}}, \bibinfo {author} {\bibfnamefont {T.}~\bibnamefont {Kato}},\ and\ \bibinfo {author} {\bibfnamefont {M.}~\bibnamefont {Ogata}},\ }\bibfield  {title} {\bibinfo {title} {{Stripe states with spatially oscillating d-wave superconductivity in the two-dimensional t-t'-J model}},\ }\href {https://doi.org/10.1103/PhysRevLett.88.117001} {\bibfield  {journal} {\bibinfo  {journal} {Physical review letters}\ }\textbf {\bibinfo {volume} {88}},\ \bibinfo {pages} {117001} (\bibinfo {year} {2002})}\BibitemShut {NoStop}%
\bibitem [{\citenamefont {Corboz}\ \emph {et~al.}(2014)\citenamefont {Corboz}, \citenamefont {Rice},\ and\ \citenamefont {Troyer}}]{Corboz2014-kr}%
  \BibitemOpen
  \bibfield  {author} {\bibinfo {author} {\bibfnamefont {P.}~\bibnamefont {Corboz}}, \bibinfo {author} {\bibfnamefont {T.~M.}\ \bibnamefont {Rice}},\ and\ \bibinfo {author} {\bibfnamefont {M.}~\bibnamefont {Troyer}},\ }\bibfield  {title} {\bibinfo {title} {{Competing states in the t-J model: uniform D-wave state versus stripe state}},\ }\href {https://doi.org/10.1103/PhysRevLett.113.046402} {\bibfield  {journal} {\bibinfo  {journal} {Physical review letters}\ }\textbf {\bibinfo {volume} {113}},\ \bibinfo {pages} {046402} (\bibinfo {year} {2014})}\BibitemShut {NoStop}%
\bibitem [{\citenamefont {Zheng}\ and\ \citenamefont {Chan}(2016)}]{Zheng2016-xt}%
  \BibitemOpen
  \bibfield  {author} {\bibinfo {author} {\bibfnamefont {B.-X.}\ \bibnamefont {Zheng}}\ and\ \bibinfo {author} {\bibfnamefont {G.~K.-L.}\ \bibnamefont {Chan}},\ }\bibfield  {title} {\bibinfo {title} {{Ground-state phase diagram of the square lattice Hubbard model from density matrix embedding theory}},\ }\href {https://doi.org/10.1103/PhysRevB.93.035126} {\bibfield  {journal} {\bibinfo  {journal} {Physical review. B, Condensed matter}\ }\textbf {\bibinfo {volume} {93}},\ \bibinfo {pages} {035126} (\bibinfo {year} {2016})}\BibitemShut {NoStop}%
\bibitem [{\citenamefont {Chen}\ \emph {et~al.}(2021)\citenamefont {Chen}, \citenamefont {Wang}, \citenamefont {Rebec}, \citenamefont {Jia}, \citenamefont {Hashimoto}, \citenamefont {Lu}, \citenamefont {Moritz}, \citenamefont {Moore}, \citenamefont {Devereaux},\ and\ \citenamefont {Shen}}]{Chen2021-xx}%
  \BibitemOpen
  \bibfield  {author} {\bibinfo {author} {\bibfnamefont {Z.}~\bibnamefont {Chen}}, \bibinfo {author} {\bibfnamefont {Y.}~\bibnamefont {Wang}}, \bibinfo {author} {\bibfnamefont {S.~N.}\ \bibnamefont {Rebec}}, \bibinfo {author} {\bibfnamefont {T.}~\bibnamefont {Jia}}, \bibinfo {author} {\bibfnamefont {M.}~\bibnamefont {Hashimoto}}, \bibinfo {author} {\bibfnamefont {D.}~\bibnamefont {Lu}}, \bibinfo {author} {\bibfnamefont {B.}~\bibnamefont {Moritz}}, \bibinfo {author} {\bibfnamefont {R.~G.}\ \bibnamefont {Moore}}, \bibinfo {author} {\bibfnamefont {T.~P.}\ \bibnamefont {Devereaux}},\ and\ \bibinfo {author} {\bibfnamefont {Z.-X.}\ \bibnamefont {Shen}},\ }\bibfield  {title} {\bibinfo {title} {{Anomalously strong near-neighbor attraction in doped 1D cuprate chains}},\ }\href {https://doi.org/10.1126/science.abf5174} {\bibfield  {journal} {\bibinfo  {journal} {Science}\ }\textbf {\bibinfo {volume} {373}},\ \bibinfo {pages} {1235} (\bibinfo {year} {2021})}\BibitemShut {NoStop}%
\bibitem [{\citenamefont {Wang}\ \emph {et~al.}(2021)\citenamefont {Wang}, \citenamefont {Chen}, \citenamefont {Shi}, \citenamefont {Moritz}, \citenamefont {Shen},\ and\ \citenamefont {Devereaux}}]{Wang2021-vs}%
  \BibitemOpen
  \bibfield  {author} {\bibinfo {author} {\bibfnamefont {Y.}~\bibnamefont {Wang}}, \bibinfo {author} {\bibfnamefont {Z.}~\bibnamefont {Chen}}, \bibinfo {author} {\bibfnamefont {T.}~\bibnamefont {Shi}}, \bibinfo {author} {\bibfnamefont {B.}~\bibnamefont {Moritz}}, \bibinfo {author} {\bibfnamefont {Z.-X.}\ \bibnamefont {Shen}},\ and\ \bibinfo {author} {\bibfnamefont {T.~P.}\ \bibnamefont {Devereaux}},\ }\bibfield  {title} {\bibinfo {title} {{Phonon-Mediated Long-Range Attractive Interaction in One-Dimensional Cuprates}},\ }\href {https://doi.org/10.1103/PhysRevLett.127.197003} {\bibfield  {journal} {\bibinfo  {journal} {Physical review letters}\ }\textbf {\bibinfo {volume} {127}},\ \bibinfo {pages} {197003} (\bibinfo {year} {2021})}\BibitemShut {NoStop}%
\bibitem [{\citenamefont {Tang}\ \emph {et~al.}(2023)\citenamefont {Tang}, \citenamefont {Moritz}, \citenamefont {Peng}, \citenamefont {Shen},\ and\ \citenamefont {Devereaux}}]{Tang2023-aa}%
  \BibitemOpen
  \bibfield  {author} {\bibinfo {author} {\bibfnamefont {T.}~\bibnamefont {Tang}}, \bibinfo {author} {\bibfnamefont {B.}~\bibnamefont {Moritz}}, \bibinfo {author} {\bibfnamefont {C.}~\bibnamefont {Peng}}, \bibinfo {author} {\bibfnamefont {Z.-X.}\ \bibnamefont {Shen}},\ and\ \bibinfo {author} {\bibfnamefont {T.~P.}\ \bibnamefont {Devereaux}},\ }\bibfield  {title} {\bibinfo {title} {{Traces of electron-phonon coupling in one-dimensional cuprates}},\ }\href {https://doi.org/10.1038/s41467-023-38408-6} {\bibfield  {journal} {\bibinfo  {journal} {Nature communications}\ }\textbf {\bibinfo {volume} {14}},\ \bibinfo {pages} {3129} (\bibinfo {year} {2023})}\BibitemShut {NoStop}%
\bibitem [{\citenamefont {Peng}\ \emph {et~al.}(2023)\citenamefont {Peng}, \citenamefont {Wang}, \citenamefont {Wen}, \citenamefont {Lee}, \citenamefont {Devereaux},\ and\ \citenamefont {Jiang}}]{Peng2023-on}%
  \BibitemOpen
  \bibfield  {author} {\bibinfo {author} {\bibfnamefont {C.}~\bibnamefont {Peng}}, \bibinfo {author} {\bibfnamefont {Y.}~\bibnamefont {Wang}}, \bibinfo {author} {\bibfnamefont {J.}~\bibnamefont {Wen}}, \bibinfo {author} {\bibfnamefont {Y.~S.}\ \bibnamefont {Lee}}, \bibinfo {author} {\bibfnamefont {T.~P.}\ \bibnamefont {Devereaux}},\ and\ \bibinfo {author} {\bibfnamefont {H.-C.}\ \bibnamefont {Jiang}},\ }\bibfield  {title} {\bibinfo {title} {{Enhanced superconductivity by near-neighbor attraction in the doped extended Hubbard model}},\ }\href {https://doi.org/10.1103/PhysRevB.107.L201102} {\bibfield  {journal} {\bibinfo  {journal} {Physical review. B, Condensed matter}\ }\textbf {\bibinfo {volume} {107}},\ \bibinfo {pages} {L201102} (\bibinfo {year} {2023})}\BibitemShut {NoStop}%
\bibitem [{\citenamefont {Cao}\ \emph {et~al.}(2025)\citenamefont {Cao}, \citenamefont {Li}, \citenamefont {Su}, \citenamefont {Ying},\ and\ \citenamefont {Tang}}]{Cao2024-hj}%
  \BibitemOpen
  \bibfield  {author} {\bibinfo {author} {\bibfnamefont {Z.}~\bibnamefont {Cao}}, \bibinfo {author} {\bibfnamefont {J.}~\bibnamefont {Li}}, \bibinfo {author} {\bibfnamefont {J.}~\bibnamefont {Su}}, \bibinfo {author} {\bibfnamefont {T.}~\bibnamefont {Ying}},\ and\ \bibinfo {author} {\bibfnamefont {H.-K.}\ \bibnamefont {Tang}},\ }\bibfield  {title} {\bibinfo {title} {Dominant $p$-wave pairing induced by nearest-neighbor attraction in the square-lattice extended hubbard model},\ }\href {https://doi.org/10.1103/PhysRevB.111.024509} {\bibfield  {journal} {\bibinfo  {journal} {Phys. Rev. B}\ }\textbf {\bibinfo {volume} {111}},\ \bibinfo {pages} {024509} (\bibinfo {year} {2025})}\BibitemShut {NoStop}%
\bibitem [{\citenamefont {Lu}\ \emph {et~al.}(2024{\natexlab{b}})\citenamefont {Lu}, \citenamefont {Guo}, \citenamefont {Chen}, \citenamefont {Sheng},\ and\ \citenamefont {Gong}}]{Lu2024-bh}%
  \BibitemOpen
  \bibfield  {author} {\bibinfo {author} {\bibfnamefont {X.}~\bibnamefont {Lu}}, \bibinfo {author} {\bibfnamefont {H.}~\bibnamefont {Guo}}, \bibinfo {author} {\bibfnamefont {W.-Q.}\ \bibnamefont {Chen}}, \bibinfo {author} {\bibfnamefont {D.~N.}\ \bibnamefont {Sheng}},\ and\ \bibinfo {author} {\bibfnamefont {S.-S.}\ \bibnamefont {Gong}},\ }\bibfield  {title} {\bibinfo {title} {{Tuning competition between charge order and superconductivity in the square-lattice $t$-$t'$-$J$ model}},\ }\href@noop {} {\bibfield  {journal} {\bibinfo  {journal} {arXiv [cond-mat.str-el]}\ } (\bibinfo {year} {2024}{\natexlab{b}})},\ \Eprint {https://arxiv.org/abs/2409.15270} {arXiv:2409.15270 [cond-mat.str-el]} \BibitemShut {NoStop}%
\bibitem [{\citenamefont {Misawa}\ and\ \citenamefont {Imada}(2014)}]{Misawa2014-at}%
  \BibitemOpen
  \bibfield  {author} {\bibinfo {author} {\bibfnamefont {T.}~\bibnamefont {Misawa}}\ and\ \bibinfo {author} {\bibfnamefont {M.}~\bibnamefont {Imada}},\ }\bibfield  {title} {\bibinfo {title} {{Origin of high-Tcsuperconductivity in doped Hubbard models and their extensions: Roles of uniform charge fluctuations}},\ }\bibfield  {journal} {\bibinfo  {journal} {Physical review. B, Condensed matter and materials physics}\ }\textbf {\bibinfo {volume} {90}},\ \href {https://doi.org/10.1103/physrevb.90.115137} {10.1103/physrevb.90.115137} (\bibinfo {year} {2014})\BibitemShut {NoStop}%
\bibitem [{\citenamefont {Sau}\ and\ \citenamefont {Sachdev}(2014)}]{Sau2014-pc}%
  \BibitemOpen
  \bibfield  {author} {\bibinfo {author} {\bibfnamefont {J.~D.}\ \bibnamefont {Sau}}\ and\ \bibinfo {author} {\bibfnamefont {S.}~\bibnamefont {Sachdev}},\ }\bibfield  {title} {\bibinfo {title} {{Mean-field theory of competing orders in metals with antiferromagnetic exchange interactions}},\ }\bibfield  {journal} {\bibinfo  {journal} {Physical review. B, Condensed matter and materials physics}\ }\textbf {\bibinfo {volume} {89}},\ \href {https://doi.org/10.1103/physrevb.89.075129} {10.1103/physrevb.89.075129} (\bibinfo {year} {2014})\BibitemShut {NoStop}%
\bibitem [{\citenamefont {Jiang}\ \emph {et~al.}(2018)\citenamefont {Jiang}, \citenamefont {H{\"{a}}hner}, \citenamefont {Schulthess},\ and\ \citenamefont {Maier}}]{Jiang2018-sc}%
  \BibitemOpen
  \bibfield  {author} {\bibinfo {author} {\bibfnamefont {M.}~\bibnamefont {Jiang}}, \bibinfo {author} {\bibfnamefont {U.~R.}\ \bibnamefont {H{\"{a}}hner}}, \bibinfo {author} {\bibfnamefont {T.~C.}\ \bibnamefont {Schulthess}},\ and\ \bibinfo {author} {\bibfnamefont {T.~A.}\ \bibnamefont {Maier}},\ }\bibfield  {title} {\bibinfo {title} {{d -wave superconductivity in the presence of nearest-neighbor Coulomb repulsion}},\ }\bibfield  {journal} {\bibinfo  {journal} {Physical review. B}\ }\textbf {\bibinfo {volume} {97}},\ \href {https://doi.org/10.1103/physrevb.97.184507} {10.1103/physrevb.97.184507} (\bibinfo {year} {2018})\BibitemShut {NoStop}%
\bibitem [{\citenamefont {Hirayama}\ \emph {et~al.}(2018)\citenamefont {Hirayama}, \citenamefont {Yamaji}, \citenamefont {Misawa},\ and\ \citenamefont {Imada}}]{Hirayama2018-tm}%
  \BibitemOpen
  \bibfield  {author} {\bibinfo {author} {\bibfnamefont {M.}~\bibnamefont {Hirayama}}, \bibinfo {author} {\bibfnamefont {Y.}~\bibnamefont {Yamaji}}, \bibinfo {author} {\bibfnamefont {T.}~\bibnamefont {Misawa}},\ and\ \bibinfo {author} {\bibfnamefont {M.}~\bibnamefont {Imada}},\ }\bibfield  {title} {\bibinfo {title} {{Ab initio effective Hamiltonians for cuprate superconductors}},\ }\href {https://doi.org/10.1103/PhysRevB.98.134501} {\bibfield  {journal} {\bibinfo  {journal} {Physical review. B, Condensed matter}\ }\textbf {\bibinfo {volume} {98}},\ \bibinfo {pages} {134501} (\bibinfo {year} {2018})}\BibitemShut {NoStop}%
\bibitem [{\citenamefont {Banerjee}\ \emph {et~al.}(2022)\citenamefont {Banerjee}, \citenamefont {P\'{e}pin},\ and\ \citenamefont {Ghosal}}]{Banerjee2022-vp}%
  \BibitemOpen
  \bibfield  {author} {\bibinfo {author} {\bibfnamefont {A.}~\bibnamefont {Banerjee}}, \bibinfo {author} {\bibfnamefont {C.}~\bibnamefont {P\'{e}pin}},\ and\ \bibinfo {author} {\bibfnamefont {A.}~\bibnamefont {Ghosal}},\ }\bibfield  {title} {\bibinfo {title} {{Charge, bond, and pair density wave orders in a strongly correlated system}},\ }\bibfield  {journal} {\bibinfo  {journal} {Physical review. B}\ }\textbf {\bibinfo {volume} {105}},\ \href {https://doi.org/10.1103/physrevb.105.134505} {10.1103/physrevb.105.134505} (\bibinfo {year} {2022})\BibitemShut {NoStop}%
\bibitem [{\citenamefont {Bejas}\ \emph {et~al.}(2022)\citenamefont {Bejas}, \citenamefont {Zeyher},\ and\ \citenamefont {Greco}}]{Bejas2022-gk}%
  \BibitemOpen
  \bibfield  {author} {\bibinfo {author} {\bibfnamefont {M.}~\bibnamefont {Bejas}}, \bibinfo {author} {\bibfnamefont {R.}~\bibnamefont {Zeyher}},\ and\ \bibinfo {author} {\bibfnamefont {A.}~\bibnamefont {Greco}},\ }\bibfield  {title} {\bibinfo {title} {{Ring-like shaped charge modulations in the t-J model with long-range Coulomb interaction}},\ }\bibfield  {journal} {\bibinfo  {journal} {Physical review. B}\ }\textbf {\bibinfo {volume} {106}},\ \href {https://doi.org/10.1103/physrevb.106.224512} {10.1103/physrevb.106.224512} (\bibinfo {year} {2022})\BibitemShut {NoStop}%
\bibitem [{\citenamefont {Boschini}\ \emph {et~al.}(2021)\citenamefont {Boschini}, \citenamefont {Minola}, \citenamefont {Sutarto}, \citenamefont {Schierle}, \citenamefont {Bluschke}, \citenamefont {Das}, \citenamefont {Yang}, \citenamefont {Michiardi}, \citenamefont {Shao}, \citenamefont {Feng}, \citenamefont {Ono}, \citenamefont {Zhong}, \citenamefont {Schneeloch}, \citenamefont {Gu}, \citenamefont {Weschke}, \citenamefont {He}, \citenamefont {Chuang}, \citenamefont {Keimer}, \citenamefont {Damascelli}, \citenamefont {Frano},\ and\ \citenamefont {da~Silva~Neto}}]{Boschini2021-xf}%
  \BibitemOpen
  \bibfield  {author} {\bibinfo {author} {\bibfnamefont {F.}~\bibnamefont {Boschini}}, \bibinfo {author} {\bibfnamefont {M.}~\bibnamefont {Minola}}, \bibinfo {author} {\bibfnamefont {R.}~\bibnamefont {Sutarto}}, \bibinfo {author} {\bibfnamefont {E.}~\bibnamefont {Schierle}}, \bibinfo {author} {\bibfnamefont {M.}~\bibnamefont {Bluschke}}, \bibinfo {author} {\bibfnamefont {S.}~\bibnamefont {Das}}, \bibinfo {author} {\bibfnamefont {Y.}~\bibnamefont {Yang}}, \bibinfo {author} {\bibfnamefont {M.}~\bibnamefont {Michiardi}}, \bibinfo {author} {\bibfnamefont {Y.~C.}\ \bibnamefont {Shao}}, \bibinfo {author} {\bibfnamefont {X.}~\bibnamefont {Feng}}, \bibinfo {author} {\bibfnamefont {S.}~\bibnamefont {Ono}}, \bibinfo {author} {\bibfnamefont {R.~D.}\ \bibnamefont {Zhong}}, \bibinfo {author} {\bibfnamefont {J.~A.}\ \bibnamefont {Schneeloch}}, \bibinfo {author} {\bibfnamefont {G.~D.}\ \bibnamefont {Gu}}, \bibinfo {author} {\bibfnamefont {E.}~\bibnamefont {Weschke}}, \bibinfo {author} {\bibfnamefont {F.}~\bibnamefont {He}},
  \bibinfo {author} {\bibfnamefont {Y.~D.}\ \bibnamefont {Chuang}}, \bibinfo {author} {\bibfnamefont {B.}~\bibnamefont {Keimer}}, \bibinfo {author} {\bibfnamefont {A.}~\bibnamefont {Damascelli}}, \bibinfo {author} {\bibfnamefont {A.}~\bibnamefont {Frano}},\ and\ \bibinfo {author} {\bibfnamefont {E.~H.}\ \bibnamefont {da~Silva~Neto}},\ }\bibfield  {title} {\bibinfo {title} {{Dynamic electron correlations with charge order wavelength along all directions in the copper oxide plane}},\ }\href {https://doi.org/10.1038/s41467-020-20824-7} {\bibfield  {journal} {\bibinfo  {journal} {Nature communications}\ }\textbf {\bibinfo {volume} {12}},\ \bibinfo {pages} {1} (\bibinfo {year} {2021})}\BibitemShut {NoStop}%
\bibitem [{\citenamefont {S\'en\'echal}\ \emph {et~al.}(2013)\citenamefont {S\'en\'echal}, \citenamefont {Day}, \citenamefont {Bouliane},\ and\ \citenamefont {Tremblay}}]{PhysRevB.87.075123}%
  \BibitemOpen
  \bibfield  {author} {\bibinfo {author} {\bibfnamefont {D.}~\bibnamefont {S\'en\'echal}}, \bibinfo {author} {\bibfnamefont {A.~G.~R.}\ \bibnamefont {Day}}, \bibinfo {author} {\bibfnamefont {V.}~\bibnamefont {Bouliane}},\ and\ \bibinfo {author} {\bibfnamefont {A.-M.~S.}\ \bibnamefont {Tremblay}},\ }\bibfield  {title} {\bibinfo {title} {Resilience of $d$-wave superconductivity to nearest-neighbor repulsion},\ }\href {https://doi.org/10.1103/PhysRevB.87.075123} {\bibfield  {journal} {\bibinfo  {journal} {Phys. Rev. B}\ }\textbf {\bibinfo {volume} {87}},\ \bibinfo {pages} {075123} (\bibinfo {year} {2013})}\BibitemShut {NoStop}%
\bibitem [{\citenamefont {Reymbaut}\ \emph {et~al.}(2016)\citenamefont {Reymbaut}, \citenamefont {Charlebois}, \citenamefont {Asiani}, \citenamefont {Fratino}, \citenamefont {S\'emon}, \citenamefont {Sordi},\ and\ \citenamefont {Tremblay}}]{PhysRevB.94.155146}%
  \BibitemOpen
  \bibfield  {author} {\bibinfo {author} {\bibfnamefont {A.}~\bibnamefont {Reymbaut}}, \bibinfo {author} {\bibfnamefont {M.}~\bibnamefont {Charlebois}}, \bibinfo {author} {\bibfnamefont {M.~F.}\ \bibnamefont {Asiani}}, \bibinfo {author} {\bibfnamefont {L.}~\bibnamefont {Fratino}}, \bibinfo {author} {\bibfnamefont {P.}~\bibnamefont {S\'emon}}, \bibinfo {author} {\bibfnamefont {G.}~\bibnamefont {Sordi}},\ and\ \bibinfo {author} {\bibfnamefont {A.-M.~S.}\ \bibnamefont {Tremblay}},\ }\bibfield  {title} {\bibinfo {title} {Antagonistic effects of nearest-neighbor repulsion on the superconducting pairing dynamics in the doped mott insulator regime},\ }\href {https://doi.org/10.1103/PhysRevB.94.155146} {\bibfield  {journal} {\bibinfo  {journal} {Phys. Rev. B}\ }\textbf {\bibinfo {volume} {94}},\ \bibinfo {pages} {155146} (\bibinfo {year} {2016})}\BibitemShut {NoStop}%
\bibitem [{\citenamefont {Jiang}(2022)}]{Jiang2022-ok}%
  \BibitemOpen
  \bibfield  {author} {\bibinfo {author} {\bibfnamefont {M.}~\bibnamefont {Jiang}},\ }\bibfield  {title} {\bibinfo {title} {{Enhancing $d$-wave superconductivity with nearest-neighbor attraction in the extended Hubbard model}},\ }\href {https://doi.org/10.1103/PhysRevB.105.024510} {\bibfield  {journal} {\bibinfo  {journal} {Physical review. B, Condensed matter}\ }\textbf {\bibinfo {volume} {105}},\ \bibinfo {pages} {024510} (\bibinfo {year} {2022})}\BibitemShut {NoStop}%
\bibitem [{\citenamefont {Andersen}\ \emph {et~al.}(1995)\citenamefont {Andersen}, \citenamefont {Liechtenstein}, \citenamefont {Jepsen},\ and\ \citenamefont {Paulsen}}]{Andersen1995-pi}%
  \BibitemOpen
  \bibfield  {author} {\bibinfo {author} {\bibfnamefont {O.~K.}\ \bibnamefont {Andersen}}, \bibinfo {author} {\bibfnamefont {A.~I.}\ \bibnamefont {Liechtenstein}}, \bibinfo {author} {\bibfnamefont {O.}~\bibnamefont {Jepsen}},\ and\ \bibinfo {author} {\bibfnamefont {F.}~\bibnamefont {Paulsen}},\ }\bibfield  {title} {\bibinfo {title} {{LDA energy bands, low-energy hamiltonians, $t^{\prime}$, $t^{\prime\prime}$, $t_{\bot}(k)$, and $J_{\bot}$}},\ }\href {https://doi.org/10.1016/0022-3697(95)00269-3} {\bibfield  {journal} {\bibinfo  {journal} {The Journal of physics and chemistry of solids}\ }\textbf {\bibinfo {volume} {56}},\ \bibinfo {pages} {1573} (\bibinfo {year} {1995})}\BibitemShut {NoStop}%
\bibitem [{\citenamefont {Zhang}\ \emph {et~al.}(1995)\citenamefont {Zhang}, \citenamefont {Carlson},\ and\ \citenamefont {Gubernatis}}]{Zhang1995-hn}%
  \BibitemOpen
  \bibfield  {author} {\bibinfo {author} {\bibfnamefont {S.}~\bibnamefont {Zhang}}, \bibinfo {author} {\bibfnamefont {J.}~\bibnamefont {Carlson}},\ and\ \bibinfo {author} {\bibfnamefont {J.~E.}\ \bibnamefont {Gubernatis}},\ }\bibfield  {title} {\bibinfo {title} {{Constrained path quantum Monte Carlo method for fermion ground states}},\ }\href {https://doi.org/10.1103/PhysRevLett.74.3652} {\bibfield  {journal} {\bibinfo  {journal} {Physical review letters}\ }\textbf {\bibinfo {volume} {74}},\ \bibinfo {pages} {3652} (\bibinfo {year} {1995})}\BibitemShut {NoStop}%
\bibitem [{\citenamefont {Zhang}\ \emph {et~al.}(1997)\citenamefont {Zhang}, \citenamefont {Carlson},\ and\ \citenamefont {Gubernatis}}]{Zhang1997-mr}%
  \BibitemOpen
  \bibfield  {author} {\bibinfo {author} {\bibfnamefont {S.}~\bibnamefont {Zhang}}, \bibinfo {author} {\bibfnamefont {J.}~\bibnamefont {Carlson}},\ and\ \bibinfo {author} {\bibfnamefont {J.~E.}\ \bibnamefont {Gubernatis}},\ }\bibfield  {title} {\bibinfo {title} {{Constrained path Monte Carlo method for fermion ground states}},\ }\href {https://doi.org/10.1103/PhysRevB.55.7464} {\bibfield  {journal} {\bibinfo  {journal} {Physical review. B, Condensed matter}\ }\textbf {\bibinfo {volume} {55}},\ \bibinfo {pages} {7464} (\bibinfo {year} {1997})}\BibitemShut {NoStop}%
\bibitem [{\citenamefont {White}\ \emph {et~al.}(1989)\citenamefont {White}, \citenamefont {Scalapino}, \citenamefont {Sugar}, \citenamefont {Bickers},\ and\ \citenamefont {Scalettar}}]{White1989-fn}%
  \BibitemOpen
  \bibfield  {author} {\bibinfo {author} {\bibfnamefont {S.~R.}\ \bibnamefont {White}}, \bibinfo {author} {\bibfnamefont {D.~J.}\ \bibnamefont {Scalapino}}, \bibinfo {author} {\bibfnamefont {R.~L.}\ \bibnamefont {Sugar}}, \bibinfo {author} {\bibfnamefont {N.~E.}\ \bibnamefont {Bickers}},\ and\ \bibinfo {author} {\bibfnamefont {R.~T.}\ \bibnamefont {Scalettar}},\ }\bibfield  {title} {\bibinfo {title} {{Attractive and repulsive pairing interaction vertices for the two-dimensional Hubbard model}},\ }\href {https://doi.org/10.1103/physrevb.39.839} {\bibfield  {journal} {\bibinfo  {journal} {Physical review. B, Condensed matter}\ }\textbf {\bibinfo {volume} {39}},\ \bibinfo {pages} {839} (\bibinfo {year} {1989})}\BibitemShut {NoStop}%
\bibitem [{\citenamefont {Ying}\ and\ \citenamefont {Wessel}(2018)}]{Ying2018-ev}%
  \BibitemOpen
  \bibfield  {author} {\bibinfo {author} {\bibfnamefont {T.}~\bibnamefont {Ying}}\ and\ \bibinfo {author} {\bibfnamefont {S.}~\bibnamefont {Wessel}},\ }\bibfield  {title} {\bibinfo {title} {{Pairing and chiral spin density wave instabilities on the honeycomb lattice: A comparative quantum Monte Carlo study}},\ }\bibfield  {journal} {\bibinfo  {journal} {Physical review. B}\ }\textbf {\bibinfo {volume} {97}},\ \href {https://doi.org/10.1103/physrevb.97.075127} {10.1103/physrevb.97.075127} (\bibinfo {year} {2018})\BibitemShut {NoStop}%
\bibitem [{\citenamefont {Dahm}\ \emph {et~al.}(2009)\citenamefont {Dahm}, \citenamefont {Hinkov}, \citenamefont {Borisenko}, \citenamefont {Kordyuk}, \citenamefont {Zabolotnyy}, \citenamefont {Fink}, \citenamefont {Büchner}, \citenamefont {Scalapino}, \citenamefont {Hanke},\ and\ \citenamefont {Keimer}}]{Dahm2009-wg}%
  \BibitemOpen
  \bibfield  {author} {\bibinfo {author} {\bibfnamefont {T.}~\bibnamefont {Dahm}}, \bibinfo {author} {\bibfnamefont {V.}~\bibnamefont {Hinkov}}, \bibinfo {author} {\bibfnamefont {S.~V.}\ \bibnamefont {Borisenko}}, \bibinfo {author} {\bibfnamefont {A.~A.}\ \bibnamefont {Kordyuk}}, \bibinfo {author} {\bibfnamefont {V.~B.}\ \bibnamefont {Zabolotnyy}}, \bibinfo {author} {\bibfnamefont {J.}~\bibnamefont {Fink}}, \bibinfo {author} {\bibfnamefont {B.}~\bibnamefont {Büchner}}, \bibinfo {author} {\bibfnamefont {D.~J.}\ \bibnamefont {Scalapino}}, \bibinfo {author} {\bibfnamefont {W.}~\bibnamefont {Hanke}},\ and\ \bibinfo {author} {\bibfnamefont {B.}~\bibnamefont {Keimer}},\ }\bibfield  {title} {\bibinfo {title} {{Strength of the spin-fluctuation-mediated pairing interaction in a high-temperature superconductor}},\ }\href {https://doi.org/10.1038/nphys1180} {\bibfield  {journal} {\bibinfo  {journal} {Nature physics}\ }\textbf {\bibinfo {volume} {5}},\ \bibinfo {pages} {217} (\bibinfo {year} {2009})}\BibitemShut
  {NoStop}%
\bibitem [{\citenamefont {Dong}\ \emph {et~al.}(2022)\citenamefont {Dong}, \citenamefont {Del~Re}, \citenamefont {Toschi},\ and\ \citenamefont {Gull}}]{Dong2022-xh}%
  \BibitemOpen
  \bibfield  {author} {\bibinfo {author} {\bibfnamefont {X.}~\bibnamefont {Dong}}, \bibinfo {author} {\bibfnamefont {L.}~\bibnamefont {Del~Re}}, \bibinfo {author} {\bibfnamefont {A.}~\bibnamefont {Toschi}},\ and\ \bibinfo {author} {\bibfnamefont {E.}~\bibnamefont {Gull}},\ }\bibfield  {title} {\bibinfo {title} {{Mechanism of superconductivity in the Hubbard model at intermediate interaction strength}},\ }\href {https://doi.org/10.1073/pnas.2205048119} {\bibfield  {journal} {\bibinfo  {journal} {Proceedings of the National Academy of Sciences of the United States of America}\ }\textbf {\bibinfo {volume} {119}},\ \bibinfo {pages} {e2205048119} (\bibinfo {year} {2022})}\BibitemShut {NoStop}%
\bibitem [{\citenamefont {Chang}\ and\ \citenamefont {Zhang}(2010)}]{Chang2010-za}%
  \BibitemOpen
  \bibfield  {author} {\bibinfo {author} {\bibfnamefont {C.-C.}\ \bibnamefont {Chang}}\ and\ \bibinfo {author} {\bibfnamefont {S.}~\bibnamefont {Zhang}},\ }\bibfield  {title} {\bibinfo {title} {{Spin and charge order in the doped hubbard model: long-wavelength collective modes}},\ }\href {https://doi.org/10.1103/PhysRevLett.104.116402} {\bibfield  {journal} {\bibinfo  {journal} {Physical review letters}\ }\textbf {\bibinfo {volume} {104}},\ \bibinfo {pages} {116402} (\bibinfo {year} {2010})}\BibitemShut {NoStop}%
\bibitem [{\citenamefont {Zhang}\ \emph {et~al.}(2022)\citenamefont {Zhang}, \citenamefont {Guo}, \citenamefont {Mou}, \citenamefont {Chen},\ and\ \citenamefont {Ma}}]{Zhang2022-ej}%
  \BibitemOpen
  \bibfield  {author} {\bibinfo {author} {\bibfnamefont {L.}~\bibnamefont {Zhang}}, \bibinfo {author} {\bibfnamefont {T.}~\bibnamefont {Guo}}, \bibinfo {author} {\bibfnamefont {Y.}~\bibnamefont {Mou}}, \bibinfo {author} {\bibfnamefont {Q.}~\bibnamefont {Chen}},\ and\ \bibinfo {author} {\bibfnamefont {T.}~\bibnamefont {Ma}},\ }\bibfield  {title} {\bibinfo {title} {{Enhancement of $d$-wave pairing in the striped phase with nearest neighbor attraction}},\ }\href {https://doi.org/10.1103/PhysRevB.105.155154} {\bibfield  {journal} {\bibinfo  {journal} {Physical review. B, Condensed matter}\ }\textbf {\bibinfo {volume} {105}},\ \bibinfo {pages} {155154} (\bibinfo {year} {2022})}\BibitemShut {NoStop}%
\bibitem [{\citenamefont {Song}\ \emph {et~al.}(2017)\citenamefont {Song}, \citenamefont {Han}, \citenamefont {Kyung}, \citenamefont {Seo}, \citenamefont {Cho}, \citenamefont {Kim}, \citenamefont {Arita}, \citenamefont {Shimada}, \citenamefont {Namatame}, \citenamefont {Taniguchi}, \citenamefont {Yoshida}, \citenamefont {Eisaki}, \citenamefont {Park},\ and\ \citenamefont {Kim}}]{Song2017-bh}%
  \BibitemOpen
  \bibfield  {author} {\bibinfo {author} {\bibfnamefont {D.}~\bibnamefont {Song}}, \bibinfo {author} {\bibfnamefont {G.}~\bibnamefont {Han}}, \bibinfo {author} {\bibfnamefont {W.}~\bibnamefont {Kyung}}, \bibinfo {author} {\bibfnamefont {J.}~\bibnamefont {Seo}}, \bibinfo {author} {\bibfnamefont {S.}~\bibnamefont {Cho}}, \bibinfo {author} {\bibfnamefont {B.~S.}\ \bibnamefont {Kim}}, \bibinfo {author} {\bibfnamefont {M.}~\bibnamefont {Arita}}, \bibinfo {author} {\bibfnamefont {K.}~\bibnamefont {Shimada}}, \bibinfo {author} {\bibfnamefont {H.}~\bibnamefont {Namatame}}, \bibinfo {author} {\bibfnamefont {M.}~\bibnamefont {Taniguchi}}, \bibinfo {author} {\bibfnamefont {Y.}~\bibnamefont {Yoshida}}, \bibinfo {author} {\bibfnamefont {H.}~\bibnamefont {Eisaki}}, \bibinfo {author} {\bibfnamefont {S.~R.}\ \bibnamefont {Park}},\ and\ \bibinfo {author} {\bibfnamefont {C.}~\bibnamefont {Kim}},\ }\bibfield  {title} {\bibinfo {title} {{Electron number-based phase diagram of Pr\_\{1-x\}LaCe\_\{x\}CuO\_\{4-$\delta$\} and
  possible absence of disparity between electron- and hole-doped cuprate phase diagrams}},\ }\href {https://doi.org/10.1103/PhysRevLett.118.137001} {\bibfield  {journal} {\bibinfo  {journal} {Physical review letters}\ }\textbf {\bibinfo {volume} {118}},\ \bibinfo {pages} {137001} (\bibinfo {year} {2017})}\BibitemShut {NoStop}%
\end{thebibliography}%
\end{document}